\documentclass[%
 reprint,
superscriptaddress,
 amsmath,amssymb,
 aps,
prb,
floatfix,
]{revtex4-2}
\AtBeginDocument{
  \abovedisplayskip     =0.5\abovedisplayskip
  \abovedisplayshortskip=0.5\abovedisplayshortskip
  \belowdisplayskip     =0.5\belowdisplayskip
  \belowdisplayshortskip=0.5\belowdisplayshortskip}
  
\usepackage{graphicx}
\usepackage{dcolumn}
\usepackage{bm}
\usepackage{physics,here,siunitx,color,xcolor}
\usepackage{hyperref}
\hypersetup{
	setpagesize=false,
	bookmarksnumbered=true,%
	bookmarksopen=true,%
	colorlinks=true,%
	linkcolor=blue,
	citecolor=red,
}

\newcommand{\LL}{\mathcal{L}}
\newcommand{\ef}{\mathrm{eff}}

\newcommand{\no}{\nonumber}
\newcommand{\mr}[1]{\mathrm{#1}}

\allowdisplaybreaks

\begin{document}


\title{Theory of the inverse Faraday effect in dissipative Rashba electron systems:\texorpdfstring{\\}{}
Floquet engineering perspective
}

\author{Miho Tanaka}
 \email{22nm025r@vc.ibaraki.ac.jp}
 \affiliation{%
 Department of Physics, Ibaraki University, Mito, Ibaraki 310-8512, Japan}
 
\author{Masahiro Sato}%
 \email{sato.phys@chiba-u.jp}
\affiliation{%
 Department of Physics, Ibaraki University, Mito, Ibaraki 310-8512, Japan}%
 \affiliation{%
Department of Physics, Chiba University, Chiba 263-8522, Japan}%




\date{\today}

\begin{abstract}
We theoretically study the inverse Faraday effect (IFE), i.e., photo-induced magnetization, in two-dimensional Rashba spin-orbit coupled electron systems irradiated by a circularly polarized light. 
The quantum master (Gorini-Kossakowski-Sudarshan-Lindblad) equation 
enables us to accurately compute the laser driven dynamics, taking inevitable dissipation effects into account. 
To find the universal features of laser-driven magnetization and its dynamics, we comprehensively investigate (i) the nonequilibrium steady state (NESS) driven by a continuous wave and (ii) ultrafast spin dynamics driven by short laser pulses. In the NESS (i), 
the laser-induced magnetization and its dependence of several parameters (laser frequency, laser field strength, temperature, dissipation strength, etc.) are shown to be in good agreement with the predictions from Floquet theory for dissipative systems in the high-frequency regime. 
In the case (ii), 
we focus on ferromagnetic metal states by introducing an effective magnetic field to the Rashba model as the mean field of electron-electron interaction. 
We find that a precession of the magnetic moment occurs due to the 
pulse-driven instantaneous magnetic field and the initial phase of the precession is controlled by changing the sign of light polarization. 
This is well consistent with the spin dynamics observed in experiments of laser-pulse-driven IFE. 
We discuss how the pulse-driven dynamics are captured by the Floquet theory. Our results 
provides a microscopic method to compute ultrafast dynamics in many electron systems irradiated by intense light. 
\end{abstract}

\pacs{Valid PACS appear here}

\maketitle


\section{\label{sec:1} Introduction}
Tuning various physical properties with a time periodic field is called Floquet engineering (FE). If we focus on the laser-driven systems in solids, in the practical sense, FE means controlling static or low-frequency properties of systems by applying a higher-frequency laser field~\cite{eckardt2015, eckardt2017,oka2019, Sato2021}.  
The FE is based on the Floquet theorem for differential equations.  
Applications of Floquet theorem have a long history~\cite{shirley1965,sambe1973} 
since the 19th century, 
while in the last decade, the concept and techniques of FE have widely penetrated in the fields of condensed-matter and statistical physics. 
From the experimental viewpoint, the development of laser science has stimulated studies of FE because FE is a typical non-perturbative (non-resonant) effect of the external oscillating field and hence laser or intense electromagnetic waves are very helpful to realize FE. 

Inverse Faraday effect (IFE) has long been explored in the magneto-optics field~\cite{kirilyuk2010}, 
while it could be viewed as one of the most typical FEs in solids from the modern perspective of condensed-matter physics. 
The IFE refers to the laser-driven non-equilibrium phenomenon that a magnetization or an effective magnetic field emerges 
when we apply a circularly polarized light to magnetic materials. This could also be regarded as an ultrafast angular momentum transfer from photons to electron spins in solids. 
The IFE has been predicted in the mid 20th century~\cite{lp1961,pershan1966,hertel2006}, and the basic mechanism of IFE in solid electron systems was first revealed by the theoretical work of Ref.~\cite{pershan1966}. 
Nowadays, IFE has been observed in various magnetic materials~\cite{vanderziel1965, kimel2005a,hansteen2006, satoh2010,makino2012}
and even low-frequency-laser (THz-laser) driven IFEs in magnetic insulators (spin systems) has been also studied from the microscopic viewpoint~\cite{takayoshi2014,takayoshi2014a,ikeda2020,ikeda2021}. 

Previous studies~\cite{pershan1966,berritta2016, tanaka2020, banerjee2022} tell us that a spin-orbit (SO) interaction is essential and necessary for the emergence of laser-driven magnetic field (or magnetization) in solids. 
In solid electron systems, the necessity is naturally understood because if the AC electric field of laser is assumed to be the main driving force of IFE, the field cannot directly couple to electron spin degrees of freedom 
and the SO interaction is the unique term connecting the laser field with the spins in the Hamiltonian. 
It is proved that an SO coupling (and the resulting magnetic anisotropies) is also necessary in the IFE of spin systems~\cite{takayoshi2014,takayoshi2014a,ikeda2020,ikeda2021}.

Although (as we mentioned above) the IFE has long been investigated~\cite{pershan1966, taguchi2011, battiato2014, berritta2016, tanaka2020, dannegger2021, banerjee2022, amano2022, zhang2023a}, 
its theories starting from many-electron quantum models have not been well developed. 
In particular, the analyses based on the perspective of FE have less progressed. 
The purpose of this study is to develop such a microscopic theory, taking dissipation effects into account. 
We focus on two-dimensional (2D) square-lattice Rashba SO coupled electron models irradiated by circularly polarized light~\cite{tanaka2020} as a simple but realistic stage of IFE. To extract essential or universal features of IFE, we will concentrate on two different setups in the laser-driven electron system. The first setup is the laser-driven non-equilibrium steady states (NESSs), which are realized by waiting for a long enough time from the beginning of laser application. 
The second is the ultrafast spin dynamics when the system is irradiated by a short laser pulse. In particular, we will focus on the laser-pulse driven spin precession, which has been often observed in experiments~\cite{kimel2005a,satoh2010,makino2012}. 

One should note that dissipation effects, i.e., effects of a weak interaction between the electron system and a large environment is very important and inevitable to consider both setups. In fact, the quantum state of the system is expected to approach to a NESS due to the balance between the energy injection by laser and the energy loss by dissipation. Inversely, if we continuously apply a laser to an isolated system in solids, it is usually burnt \cite{lazarides2014,dalessio2014,kuwahara2016,mori2016}. 
Furthermore, if we ignore the dissipation effect, the energy given by a laser pulse always remains in the electron system and 
it would yield non-realistic spin dynamics, e.g., a long-time spin oscillation with no relaxation. 
We stress that the laser-driven NESS can only be reached by considering the dissipation effect.

To take such dissipation effects into account, 
we will utilize the approach of the quantum master [Gorini
Kossakowski-Sudarshan-Lindblad (GKSL)]  equation~\cite{lindblad1976,gorini1976,breuer2007,alicki1987}. 
So far, many theories for FE have been developed under the assumption that the target system is approximated by an isolated quantum system. On the other hand, some theoretical methods for FE in dissipative systems have also gradually progressed~\cite{tsuji2009,sato2020a, ikeda2020,ikeda2021,grifoni1998}. 
Relying on the theories of dissipative systems, we will reveal the fundamental properties of the laser-driven NESS and laser-pulse driven precession. These two important setups of IFEs have not been analyzed well in most of previous theories from the microscopic perspective.

The paper is organized as follows. 
Section~\ref{sec:Model} is devoted to the explanation about the SO-coupled electron model and the numerical computation method based on the GKSL equation. 
Based on these instruments, we analyze 
the continuous wave (CW) driven NESS in Sec.~\ref{sec:Ness}. 
We demonstrate that the IFE in the NESS is basically captured by the Floquet theory for dissipation systems~\cite{ikeda2020,ikeda2021} in a sufficiently high-frequency regime. 
In Sec.~\ref{sec:Pulse}, we discuss IFE driven by a short laser pulse. Since a precession of spin moment has been often detected in such short-pulse IFE for magnetic materials (see Fig.~\ref{Fig:PulsePonchi}), we prepare a ferromagnetic metal state in an electron model, by introducing a mean-field Zeeman interaction. 
We show that a pulse-driven precession is well reproduced within our model and it can be understood from the Floquet-theory viewpoint. 
In Sec.~\ref{sec:Comp}, we compare our theory with some previous theoretical works for IFEs.
Finally, in Sec.~\ref{sec:Con}, we summarize our results and comment on a few issues related to IFE. 
In the Appendix, we explain the relationship between GKSL and Bloch equations, a simple extension of the Floquet formula for dissipation systems~\cite{ikeda2020,ikeda2021}, and additional numerical results.

\section{\label{sec:Model} Model and Methods} 
In this section, we define our model of a 2D SO coupled electron model, an applied circularly polarized laser, and the GKSL equation we will use.

\subsection{\label{sec:Mode-Ham}Hamiltonian}
\begin{figure}[t]
\centering     
\includegraphics[keepaspectratio,width=\linewidth]{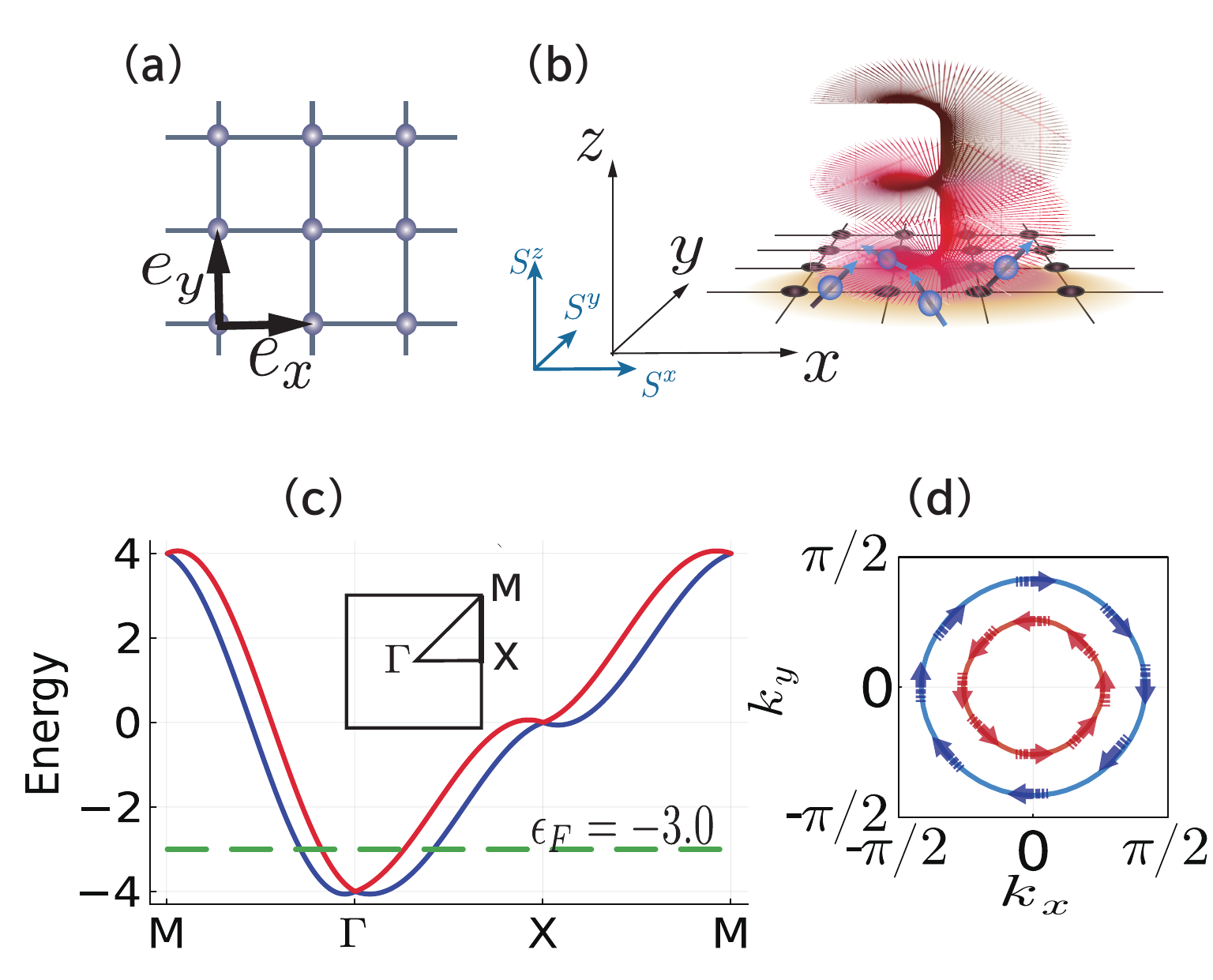}
\caption{(a) Two-dimensional square lattice. The lattice constant is set to $a=1$. Vectors $\bm e_x$ and $\bm e_y$ are respectively unit vectors connecting neighboring two sites along the $x$ and $y$ directions. 
(b) Schematic illustration of applying a circularly polarized laser whose propagating direction is perpendicular to the square lattice in the $x$--$y$ plane. 
In this setup, the effective magnetic field driven by laser (i.e., the emergent field of IFE) points to the $z$ direction. (c) Energy bands of the Hamiltonian, Eq.~(\ref{eq:hamCW}), at $t_{\mr{hop}}=$\SI{1}{eV}  and $\alpha_{\rm R}=$\SI{0.5}{eV}. Some representative points in the Brillouin zone are defined as $\Gamma=(k_x=0,k_y=0)$, $X=(k_x=\pi,k_y=0)$, and $M=(k_x=\pi,k_y=\pi)$. We set the Fermi surface at $\epsilon_F=$\SI{-3}{eV} ($t_{\mr{hop}}=$\SI{1}{eV}), which is depicted by the green dashed line. (d) Fermi surface of (c). Arrows indicate the spin orientation of electrons on the Fermi surface in the momentum space. Due to the SO coupling, a so-called spin-momentum locking occurs~\cite{e.i.rashba1960, bychkov1984,manchon2015}.
}
\label{fig:HamModel_cw}
\end{figure}
We focus on a SO-coupled tight-binding model on a 2D square lattice. 
The Hamiltonian is defined by
\begin{equation}
    \label{eq:hamCW}
    {H}_{\mr{PM}}={H}_{\mathrm{hop}}+{H}_{\mathrm{SO}}, 
\end{equation}
where we have introduced the index ``PM'' which stands for paramagnetic. 
The first term
${H}_{\mathrm{hop}}$ represents the spin-independent kinetic term, which is given by
\begin{align}
{H}_{\mathrm{hop}}=-2t_{\mr{hop}}\sum_{\bm{r},\sigma=\uparrow,\downarrow,i=x,y}
\left(c^\dagger_{\bm{r}+\bm{e}_i\sigma}c_{\bm{r}\sigma}
+ {\rm h.c.}\right),
\end{align}
where $t_{\mr{hop}}$ is the hopping strength, 
$c^\dagger_{\bm{r}\sigma}$ ($c_{\bm{r}\sigma}$) is the creation (annihilation) operator of the spin-$\sigma$ electron residing on a site $\bm r=\ell{\bm e_x}+m{\bm e_y}$ ($\ell, m\in \mathbb{Z}$) of a square lattice. These operators $c^\dagger_{\bm{r}\sigma}$ 
and $c_{\bm{r}\sigma}$ satisfy $=\{c_{\bm{r}\sigma},c^\dagger_{\bm{r}'\sigma}\}=\delta_{\bm r,\bm r'}$ and $\{c_{\bm{r}\sigma},c_{\bm{r}'\sigma}\}=\{c_{\bm{r}\sigma}^\dagger,c_{\bm{r}'\sigma}^\dagger\}=0$.
As shown in Fig.~\ref{fig:HamModel_cw}(a), 
$\bm e_x$ and $\bm e_y$ are respectively unit vectors connecting neighboring sites along the $x$ and $y$ directions and the lattice constant is set to $a=1$. 
The symbol $\sigma=\uparrow$ ($\downarrow$) corresponds to the $z$ component of electron spin $S^z=1/2$ ($-1/2$).
The second term $H_{\rm{SO}}$ is a Rashba SOI~\cite{e.i.rashba1960, bychkov1984, dresselhaus1955,manchon2015,bercioux2015}, which is defined by  
\begin{align}
   H_{\rm{SO}}=&-\frac{\alpha_{\rm R}}{2}\sum_{\bm{r}}\sum_{\sigma,\sigma'}
   \Big[\Big(i(\sigma_y)_{\sigma\sigma'}c^\dagger_{\bm{r}+\bm{e}_x,\sigma}c_{\bm{r}\sigma'}
   \nonumber\\
   &-i(\sigma_x)_{\sigma\sigma'}c^\dagger_{\bm{r}+\bm{e}_y\sigma}c_{\bm{r}\sigma'}\Big)
   +{\rm h.c.}\Big].
\end{align}
Here, $\sigma_{x,y,z}$ represents the Pauli matrices, and $\alpha_{\rm R}$ denotes the strength of Rashba SO coupling. The Pauli matrices $\sigma_{x,y}$ induce a mixing between spin-$\uparrow$ and $\downarrow$ electrons, and hence $H_{\rm{SO}}$ can be viewed as a spin-dependent hopping term. 

We define the Fourier transforms of $c^\dagger_{\bm{r}\sigma}$ and $c_{\bm{r}\sigma}$ as 
\begin{align}
c^\dagger_{\bm{r}\sigma}=\frac{1}{\sqrt{N}}\sum_{\bm{k}} e^{-i{\bm k}\cdot{\bm r}}
c^\dagger_{\bm{k}\sigma},
\label{eq:Fourier}
\end{align}
where $\bm k=(k_x,k_y)$ is the wave vector on the Brillouin zone for the 2D square lattice and $N$ is the total number of sites. 
Fermion operators $c^\dagger_{\bm{k}\sigma}$ 
and $c_{\bm{k}\sigma}$ satisfy $\{c_{\bm{k}\sigma},c^\dagger_{\bm{k}'\sigma}\}=\delta_{\bm k,\bm k'}$ and  $\{c_{\bm{k}\sigma},c_{\bm{k}'\sigma}\}=\{c_{\bm{k}\sigma}^\dagger,c_{\bm{k}'\sigma}^\dagger\}=0$. 
Substituting Eq.~(\ref{eq:Fourier}) into the Hamiltonian, we have
\begin{equation}
H_{\mathrm{hop}}=-2t_{\mr{hop}}\sum_{\bm{k}}\sum_{i=(x,y),\sigma}\cos k_i (c^\dagger_{\bm{k}\sigma}c_{\bm{k}\sigma}+{\rm h.c.}),
\end{equation}
and
\begin{align}
H_{\mr{SO}}&=\alpha_{\rm R}\sum_{\bm{k},\sigma,\sigma'}\Big\{\Big[(\sigma_y)_{\sigma\sigma'}\sin k_x\nonumber\\
     &\hspace{1.2cm}-(\sigma_x)_{\sigma\sigma'}\sin k_y\Big]c^\dagger_{\bm{k}\sigma}c_{\bm{k}\sigma'} + {\rm h.c.}\Big\}.
\end{align}
These representations show that the Hamiltonian ${H}_{\mr{PM}}$ is block diagonalized in the momentum $\bm k$ space, and one can obtain energy eigenvalues via the diagonalization in each $\bm k$ subspace as follows: 
\begin{align}              
H_{\rm PM}=\sum_{\bm{k}}
C^\dagger_{\bm k}
    \mqty(
    \varepsilon_{\bm{k}} &   \eta_{\bm k}\\
    \eta_{\bm k}^* & \varepsilon_{\bm{k}}    
    )
C_{\bm k}
=\sum_{\bm{k}}D^\dagger_{\bm k}
    \mqty(
    E^{\bm{k}}_{2}   &  0 \\
    0&   E^{\bm{k}}_{1}  
    )
    D_{\bm k},
\end{align}
where $\varepsilon_{\bm k}=-2t_{\mr{hop}}(\cos k_x+\cos k_y)$, $\eta_{\bm k}=\alpha_{\rm R}(i\sin k_x +\sin k_y)$, $C_{\bm k}={}^\mr{T}(c_{\bm{k},\uparrow},c_{\bm{k},\downarrow})$, and 
$D_{\bm k}={}^\mr{T}(d_{\bm{k},2}, d_{\bm{k},1})=U_{\bm k}C_{\bm k}$. 
The energy eigenvalues 
are given by
\begin{align}
E^{\bm k}_{1}=\varepsilon_{\bm k} - |\eta_{\bm k}|, && \;E^{\bm k}_{2}=\varepsilon_{\bm k}+ |\eta_{\bm k}|.
\label{eq:bands}
\end{align}
These values are defined such that $E^{\bm k}_{1}<E^{\bm k}_{2}$. 
The unitary matrix $U_{\bm k}$ is defined as
\begin{align}
    U_{\bm k}=\mqty(
    u_{+}^{\bm{k}} && u_{-}^{\bm{k}}\\
    v_{+}^{\bm{k}} && v_{-}^{\bm{k}})
\end{align}
with $u_{\pm}^{\bm{k}}=\frac{\eta_{\bm k}}{\sqrt{2}|\eta_{\bm k}|}$ and $v_{\pm}^{\bm{k}}=\pm \frac{1}{\sqrt{2}}$. 
The energy bands of Eq.~(\ref{eq:bands}) are shown in Fig.~\ref{fig:HamModel_cw}(c). 
In this study, the Fermi energy $\epsilon_F$ is fixed at $\epsilon_F/t_{\rm hop}=-3.0$, which is depicted in Fig.~\ref{fig:HamModel_cw}(c). The corresponding Fermi surface in the $k_x$-$k_y$ plane is given in Fig.~\ref{fig:HamModel_cw}(d). 
This parameter setup is almost the same as the previous study of IFE in Ref.~\cite{tanaka2020}. 

At the end of this subsection, we define a few spin-related operators as this work considers the laser-driven spin moments. 
The $\alpha$ component of total spin is denoted by $S^{\alpha}_{\rm{tot}}=\sum_{\bm r}S_{\bm r}^\alpha$ and those per one site are defined as 
\begin{align}
    s^\alpha=S^{\alpha}_{\rm{tot}}/N. 
\end{align}
In addition, we define the ``spin'' in the momentum space as 
\begin{align}
    s_{\bm k}^\alpha=\frac{1}{2}C_{\bm k}^\dagger \sigma_\alpha C_{\bm k}. 
\end{align}
For instance, the $z$ component of total spin is given by
\begin{align}
    S^{z}_{\rm{tot}}&=\sum_{\bm r}S_{\bm r}^z
    =\sum_{\bm r}\frac{1}{2}C^\dagger_{\bm r}\sigma_z  C_{\bm r}
    =\frac{1}{2}\sum_{\bm k}s_{\bm k}^z
\nonumber\\
    &=\frac{1}{2}\sum_{\bm k}\Big[\frac{\eta_{\bm k}}{|\eta_{\bm k}|}d_{{\bm k},2}^\dagger d_{{\bm k},1}+\frac{\eta_{\bm k}^*}{|\eta_{\bm k}|}d_{{\bm k},1}^\dagger d_{{\bm k},2}
    \Big],
\end{align}
where $C_{\bm r}={}^T(c_{\bm{r},\uparrow},c_{\bm{r},\downarrow})$. Here, we have used the unit of $\hbar=1$ and we will often use it throughout this paper.

\subsection{\label{sec:Mode-Laser}Laser}
In this study, we will consider two types of lasers. The first type is a continuous wave (CW), whose AC electric field is given by
\begin{align}  
\label{Eq:CW}
\bm{E}(t)=E_0(1-e^{-t^2/\tau_0^2})(\cos{\omega t},\sin{\omega t}, 0), 
\end{align}
where $E_0$ is the strength of electric field and $\omega$ is the angular frequency of CW laser. We make the strength of the field gradually approach to $E_0$, by introducing a parameter $\tau_0$. 
We will focus on the nonequilibrium steady states (NESSs) driven by a long-time application of CW laser and therefore $\tau_0$ is not important.   
The corresponding vector potential 
$\bm{A}(t)$ is defined as 
$-\int^{t}_{t_{\rm ini}} dt' \bm{E}(t')$ in the Coulomb gauge, where $t_{\rm ini}=0$ is the initial time. 

Another type is a laser pulse with a Gaussian wave shape. Its electric field is defined by
\begin{align}    
\label{Eq:Pulse}
\bm{E}(t)=E_0 e^{-2t^2/\tau^2}(\cos{\omega t},\sin{\omega t}, 0),
\end{align}
where $\tau$ represents the width of the pulse and we have set $\tau$ to 
$10\sqrt{2}\hbar/t_{\rm hop}\simeq 14\hbar/t_{\rm hop}$. For example, if we set $\omega=t_{\rm hop}/\hbar$, $\tau$ corresponds to about three cycles. 
The vector potential for the pulse is given by 
$\bm{A}(t)=-\int^{t}_{t_{\rm ini}} dt'\bm{E}(t')$, where the absolute value of initial time $|t_{\rm ini}|$ should be sufficiently large compared with $\tau$.  
We note that the electric fields of the above two types of lasers are spatially uniform, i.e., 
$\bm{E}(t)$ is independent of spatial coordinate $\bm r$. 
This condition is valid because the size of the laser spot is usually much larger than the lattice constant $a$ (the length scale of 1.0-0.1 nm). Typical values of the laser field, its frequency and period are summarized in Tables~\ref{Table1} and \ref{Table2} in Appendix~\ref{sec:A-Table}.

To take the effect of the AC electric field into account, we use the Peierls formalism. 
In this formalism, each hopping term $c^\dagger_{\bm{r}\sigma}c_{\bm{r}+\bm e_{i}\sigma'}$ should be replaced with $\exp[-i\frac{e}{\hbar c}\bm A(t)\cdot \bm e_{i}]c^\dagger_{\bm{r}\sigma}c_{\bm{r}+\bm e_{i}\sigma'}$ when we apply an AC electric field 
to the 2D model of Eq.~(\ref{eq:hamCW}). 
Here, $e$ $(>0)$ is the electron charge, $c$ is the speed of light, and we will often use the unit of $e=1$ and $c=1$ in this paper. 
Through the Peierls formalism, 
the time-dependent Hamiltonian of the laser-driven 2D electron model 
${H}_{\mr{PM}}(t)={H}_{\mathrm{hop}}(t)
+{H}_{\mathrm{SO}}(t)$ is given by 
\begin{align}
    H_{\mathrm{hop}}(t)=&-2t_{\mr{hop}}\sum_{\bm{k}}\sum_{i=(x,y),\sigma}\cos({A_i(t)+k_i})
    \nonumber\\
    &\hspace{3cm}\times(c^\dagger_{\bm{k}\sigma}c_{\bm{k}\sigma}+{\rm h.c.}),
    \label{eq:timeHhop}
\end{align}
and
\begin{align}
     H_{\mr{SO}} (t)&=\alpha_{\mr{R}}\sum_{\bm{k},\sigma,\sigma'}\Big\{\Big[(\sigma_y)_{\sigma\sigma'}\sin{(A_x(t)+k_x)}\nonumber\\
     &\hspace{1.2cm}-(\sigma_x)_{\sigma\sigma'}\sin{(A_y(t)+k_y})\Big]c^\dagger_{\bm{k}\sigma}c_{\bm{k}\sigma'}+ {\rm h.c.}\Big\}.
     \label{eq:timeHSO}
\end{align}
With the $2\times 2$ matrix form, we can express the Hamiltonian as  
\begin{align}
    H_{\mr{PM}}(t)=&\sum_{\bm k}H_{\mr{PM}}^{\bm k}(t),\nonumber\\
    H_{\mr{PM}}^{\bm k}(t)=&C_{\bm k}^\dagger
    \mqty(
    \varepsilon_{\bm{k},A}(t)&   \eta_{\bm{k},A}(t)\\
     \eta^*_{\bm{k},A}(t) &  \varepsilon_{\bm{k},A}(t)
    )C_{\bm k}.
    \label{ham2}
\end{align}
where $\varepsilon_{\bm{k},A}(t)$ and $\eta_{\bm{k},A}(t)$ are defined as
\begin{align}
    \varepsilon_{\bm{k},A}(t)&=-2t_{\mr{hop}}[\cos{(A_x(t)+k_x)}+\cos{(A_y(t)+k_y)}],\nonumber\\
    \eta_{\bm{k},A}(t)&=\alpha_{\mr{R}}[i\sin{(A_x(t)+k_x)} +\sin{(A_y(t)+k_y)}].
\end{align}
An important point is that even if we apply laser to the system, the time-dependent Hamiltonian, Eq.~(\ref{ham2}), is still $\bm k$-diagonal.

\subsection{\label{sec:Mode-GKSL}GKSL Equation}
To analyze the laser-driven quantum dynamics in 2D Rashba electron models, 
we use the GKSL equation~\cite{lindblad1976,gorini1976,breuer2007,alicki1987}, which is a Markovian equation of motion for the density matrix~\cite{kohler1997,hone2009,breuer2000,kohn2001} including dissipation effects. 
As we mentioned, our Hamiltonian of laser-driven systems is block diagonalized in  the momentum $\bm k$ space even after the application of laser [see Eq.~(\ref{ham2})]. 
Therefore, we may 
independently analyze the time evolution of density matrix at each $\bm k$ subspace. The GKSL equation for a subspace is defined as 
\begin{align}
\dv{\rho^{\bm k}}{t} &= -i[H^{\bm k}_{\mr{PM}}(t),\rho^{\bm k}] 
+\mathcal{D}^{\bm k}(\rho^{\bm k})
\label{eq:GKSL1}
\end{align}
where $\rho^{\bm k}(t)$ is the density matrix for the $\bm k$ subspace. 
The first commutator corresponds to the unitary dynamics driven by the Hamiltonian and 
the second term $\mathcal{D}^{\bm k}(\rho^{\bm k})$ represents the effect of dissipation and is given by
\begin{align}
\mathcal{D}^{\bm k}(\rho^{\bm k})&=\sum_{ij}\Gamma^{\bm k}_{ij}(L^{\bm k}_{ij}\rho^{\bm k} {L^{\bm k}_{ij}}^\dagger -\frac{1}{2}\{{L_{ij}^{\bm k}}^\dagger L^{\bm k}_{ij} ,\rho^{\bm k} \}).
\label{eq:GKSL2}
\end{align}
Here, a constant $\Gamma^{\bm k}_{ij}$ represents the strength of dissipation and $L^{\bm k}_{ij}$ refers to a jump (or Lindblad) operator, whose index $(i,j)$ denotes each relaxation process. 
The application of GKSL equation to our 2D electron models means that the dissipation dynamics is also assumed to be $\bm k$-diagonal. Complicated scattering processes would occur during relaxation processes in real materials, but we expect that various essential features of laser-driven dynamics can be captured by using the phenomenological dissipation term $\mathcal{D}^{\bm k}$ of the GKSL equation. In fact, 
recent studies~\cite{ikeda2019, sato2020,kanega2021, kanega2024} support this expectation. 

We assume that the jump operators do not change the electron number and such a relaxation process is natural in real experiments in solids.
From this assumption and the free electron Hamiltonian of Eq.~(\ref{ham2}), one sees that GKSL dynamics exists only in the one-particle subspace at each wavevector $\bm k$, whereas empty and two-particle states at $\bm k$ are invariant under the time evolution. 
Namely, only upper- or lower-band occupied states [see Fig.~\ref{fig:HamModel_cw}(c)] contribute to the laser-driven dynamics in our model. In this setup, the size of the density matrix $\rho^{\bm k}(t)$ is reduced to $2\times 2$. 
This property is very helpful to reduce the cost of numerical computation. 

In the present study, we define 
the jump operators as $L^{\bm k}_{ij}=\ket{E^{\bm k}_i}\bra{E^{\bm k}_j}$, where $\ket{E_{i}^{\bm k}}$ denotes $i$-th one-electron eigenstates of the time-independent Hamiltonian $H_{\mr{PM}}^{\bm k}|_{\bm A=0}$ and $E_{i}^{\bm k}$ is the corresponding $i$-th eigenenergy ($i=1,2$) under the condition of $E_{1}^{\bm k}<E_{2}^{\bm k}$. 
Furthermore, we determine the dissipation constants $\Gamma^{\bm k}_{ij}$ such that the GKSL equation satisfies the detailed balance, relaxing the system towards the equilibrium state of $H_{\rm{PM}}^{\bm k}|_{\bm A=0}$. 
Such constants $\Gamma^{\bm k}_{ij}$ are given by
\begin{align}
    \label{LOD}
    \Gamma^{\bm k}_{ij}=& \,\,\frac{\gamma\exp({-\beta E_i^{\bm k}})}{\exp({-\beta E_i^{\bm k}})+\exp({-\beta E_j^{\bm k}})},\;\;(i\neq j),\no\\
    \Gamma^{\bm k}_{ij}=& \,\,0,\;\; (i= j),
\end{align}
where $\gamma$ represents the $\bm k$-independent dissipation strength, $\beta=\frac{1}{k_BT}$ is the inverse temperature.
In zero temperature limit of 
$\beta\rightarrow\infty$, we find 
$\Gamma^{\bm k}_{12}=\gamma$ and $\Gamma^{\bm k}_{21}=0$. This indicates that in the two-level system at each $\bm k$ space, only the jump operator $L^{\bm k}_{12}=\ket{E^{\bm k}_{1}}\bra{E^{\bm k}_2}$ contributes to the dissipation dynamics and induced a transition from the excited state to the ground state.
We note that the off-diagonal jump operator $L^{\bm k}_{12}$ with Eq.~(\ref{LOD}) induces both longitudinal and transverse relaxation processes (see Appendix~\ref{sec:A-GKSLBloch}). 

On the other hand, the diagonal type of $L^{\bm k}_{i,i}$ controls the strength of 
transverse relaxation (dephasing). 
We should note that the system cannot reach a thermal equilibrium state if the GKSL equation has only diagonal type jump operators $L^{\bm k}_{i,i}$ (see Appendix~\ref{sec:A-GKSLBloch-Lz}). 
The present work will not argue the effects of $L^{\bm k}_{i,i}$ because from a few calculations, we have verified that $L^{\bm k}_{i,i}$ is not so important for IFE in our model. 

If we focus on our 2D Rashba free-electron model of Eq.~(\ref{ham2}) at $T=0$, the one-particle states at the initial time $t_{\rm ini}=0$ exist only in the doughnut regime between two Fermi surfaces in Fig.~\ref{fig:HamModel_cw}(d). 
Therefore, it is enough to numerically solve the GKSL equations with $\bm k$ being in the doughnut regime as we consider the laser-driven dynamics at $T=0$. 

The expectation value of any observable $A$ at time $t$ is defined by
\begin{align}
    \label{eq:rvA_t}
    \langle A\rangle_t=& \sum_{\bm k}{\rm Tr}[A \rho^{\bm k}(t)]. 
\end{align}
Using this formula, one can compute the time dependence of physical quantities in principle.

\section{\label{sec:Ness} Nonequilibrium Steady States}
In this section, we study the 2D Rashba electron model irradiated by a continuous-wave (CW) laser of Eq.~(\ref{Eq:CW}). 
The time parameter $\tau_0$ of Eq.~(\ref{Eq:CW}) is fixed to $50 \hbar/t_{\mr{hop}}$ throughout the numerical calculations in this section. The time-dependent Hamiltonian is given by ${H}_{\mr{PM}}(t)={H}_{\mathrm{hop}}(t)+{H}_{\mathrm{SO}}(t)$ [see Eqs.~(\ref{eq:timeHhop}) and (\ref{eq:timeHSO})]. 
We focus on the nonequilibrium steady state (NESS) that is realized if we take a sufficient time after applying a circularly polarized laser. 
From our setup of Fig.~\ref{fig:HamModel_cw}(b), a laser-driven magnetic field and magnetization are expected to emerge along the $z$ direction. Therefore, the $z$ component of spins, $s^z$ and $s^z_{\bm k}$, are the main targets of this section. 
To consider the properties of the NESS, it is useful to observe a time averaged expectation value of an observable $A$ as follows:
\begin{align}
    \langle\langle A\rangle\rangle = & \frac{1}{t_2-t_1}\int_{t_1}^{t_2} dt \langle A\rangle_t.
    \label{eq:time_av}
\end{align}
Here, $t_{1,2}-t_{\rm ini}$ ($t_{\rm ini}=0$ is the initial time) should be much larger than typical time scales of $\hbar/t_{\rm hop}$, $T_{\omega}=2\pi/\omega$ and $\hbar/\gamma$ if we want to observe the expectation value for the NESS. In addition, time interval $t_1-t_2$ should be sufficiently larger than the laser period $T_{\omega}$ if we focus on the time-independent laser-driven value. In the numerical computation of this section, we set $t_1=300\hbar/t_{\rm hop}$ and $t_2=400\hbar/t_{\rm hop}$. In all the numerical calculations of this paper, we divide the full Brillouin zone into $200\times 200$ points, which corresponds to $N=200\times 200$.

\subsection{\label{sec:Ness-Floquet} Floquet Theory}
As we already mentioned, when a CW laser is applied to the 2D Rashba model, a NESS arises due to the balance between the energy injection and dissipation. 
The density matrix for such a NESS can be analytically obtained through the recent Floquet theory approach~\cite{ikeda2020, ikeda2021} if we restrict ourselves to the high-frequency regime. In this subsection, 
we compute laser-driven magnetization in the NESS by employing the Floquet theory and high-frequency expansion~\cite{eckardt2015,mikami2016}.

First, we shortly explain the derivation of the density matrix for the NESS~\cite{ikeda2020}. 
Applying the high-frequency expansion to the GKSL equation in each $\bm k$ subspace, we derive the effective GKSL equation
\begin{align}
\label{LLFE}
\dv{\rho^{\bm k}}{t} =
\LL_{\rm{eff}}^{\bm k} \rho^{\bm k} &= -i[H_{\mr{\mr{FE}}}^{\bm k},\rho^{\bm k}] + \mathcal{D}^{\bm k}(\rho^{\bm k}),
\end{align}
which describes the lower-frequency dynamics than the laser frequency $\omega$. 
Here, the time-independent Floquet Hamiltonian $H_{\rm{\mr{FE}}}^{\bm k}$ is defined as 
\begin{align}
 \label{floH}
    H_{\rm{\mr{FE}}}^{\bm k} = H_0^{\bm k}+\sum_n\frac{[H_{-n}^{\bm k}, H_n^{\bm k}]}{n\omega}+\order{\omega^{-2}},
\end{align}
where the Fourier components of the Hamiltonian 
$H_n^{\bm k}=\frac{1}{T_\omega}\int_{0}^{T_\omega}dt\,\,H_{\mr{PM}}^{\bm k}(t)e^{in\omega t}$ $(n\in \mathbb{Z})$.
The $0$-th order term $H_0^{\bm k}$ is interpreted as the time-averaged Hamiltonian. 
From Eq.~(\ref{LLFE}), the density matrix of the NESS can be written as $\rho^{\bm k}(t)\underset{t\rightarrow \infty}{\rightarrow}\rho_{\mr{NESS}}^{\bm k}(t)=e^{\mathcal{G}(t)} \tilde{\rho}_{\infty}^{\bm k}$. The micro-motion operator $\mathcal{G}$ is a time-periodic function satisfying $\mathcal{G}(t)=\mathcal{G}(t+T_\omega)$, describing faster dynamics rather than the laser frequency. The remaining matrix $\tilde{\rho}_{\infty}^{\bm k}$ does not evolve in time and is defined by 
\begin{align}
    \mathcal{L}_{\ef}^{\bm k}\tilde{\rho}_{\infty}^{\bm k}=0. 
\end{align}
For a certain class of periodically driven systems, $H_0^{\bm k}=H_{\rm{PM}}^{\bm k}$ holds, in which the simple analytical formula of the density matrix $\rho_{\mr{NESS}}^{\bm k}(t)$ has been derived~\cite{ikeda2020}. On the other hand, for the case of $H_0^{\bm k}\neq H_{\rm{PM}}^{\bm k}$, we have to slightly extend the formula and consider the $\omega$ dependence of $H_0^{\bm k}$. Our model of the 2D Rashba model irradiated by laser corresponds to the latter case. After some algebra (see Appendix~\ref{sec:A-SolNESS}), 
the time-independent part of the density matrix 
$\tilde{\rho}_{\infty}^{\bm k}=\sum_{ij}\rho_{ij}^{\bm k}
\ket{E_i^{\bm k}}\bra{E_i^{\bm k}}$ 
is given by 
\begin{align}
\begin{split}
    \label{rhoINFham}
    \rho^{\bm k}_{11}&=\Gamma_{12}^{\bm k}/\gamma,\\
    \rho^{\bm k}_{22}&=\Gamma_{21}^{\bm k}/\gamma,\\
    \rho^{\bm k}_{12}&=\frac{\bra{E_1^{\bm k} }\Delta H_{\mr{FE}}^{\bm k}\ket{E_2^{\bm k}}}{(E_1^{\bm k}-E_2^{\bm k})-i\gamma/2}(\rho^{\bm k}_{11}-\rho^{\bm k}_{22}),\\
     \rho^{\bm k}_{21}&={\rho^{\bm k}_{21}}^*.
\end{split}
\end{align}
in the high-frequency regime. Here, 
$\Delta H_{\mr{FE}}^{\bm k}=H_{\mr{FE}}^{\bm k}-H_0^{\bm k}$. 

Using these results, we can generally compute observables of the NESS in an analytic way. 
From the definition of time-averaged expectation value in Eq.~(\ref{eq:time_av}), 
$\langle\langle s^z_{\bm k}\rangle\rangle$ is interpreted as ${\rm Tr}[s^z_{\bm k}\tilde{\rho}^{\bm k}_\infty]$. 
Combining this interpretation with Eq.~(\ref{rhoINFham}), we have 
\begin{align}
    \label{SzCW}
    \langle\langle s^z_{\bm k}\rangle\rangle
    =\frac{\bra{E_1^{\bm k}}\Delta H_{\rm{\mr{FE}}}^{\bm k}\ket{E_2^{\bm k}}(E_1^{\bm k}-E_2^{\bm k})}{(E_1^{\bm k}-E_2^{\bm k})^2+\gamma^2/4}(\rho^{\bm k}_{11}-\rho^{\bm k}_{22}).
\end{align}
Furthermore, from the formula of Eq.~(\ref{floH}), we can easily estimate the Floquet Hamiltonian as~\cite{tanaka2020} 
\begin{align}
H_{\rm{\mr{FE}}}^{\bm k}
=&\qty(1-\frac{E_0^2}{4\omega^2})
H^{\bm k}_{\rm{PM}} \nonumber\\
& - B_{\mr{eff}}({\bm k})\,\,
 \frac{1}{2}C_{\bm k}^\dagger \sigma_z C_{\bm k}
+\order{\omega^{-4}},
\label{eq:FH}
\end{align}
where 
\begin{align}
B_{\mathrm{eff}}({\bm k})
=2\frac{\alpha_{\mr{R}}^2E_0^2}{\omega^3}\cos(k_x)\cos(k_y).
\end{align}
From the fashion of the Floquet Hamiltonion, 
the parameter $B_{\mathrm{eff}}({\bm k})$ can be interpreted as the $\bm k$-space effective magnetic field driven by circularly polarized laser. In fact, $B_{\mathrm{eff}}({\bm k})$ changes its sign 
if the laser polarization changes from right to left handed ($\omega\to -\omega$).  
As the first term of Eq.~(\ref{eq:FH}) corresponds to $H_0^{\bm k}=\qty(1-\frac{E_0^2}{4\omega^2})H^{\bm k}_{\rm{PM}}$, the second term $\Delta H^{\bm k}_{\rm{\mr{FE}}}$ is expressed as
\begin{align}
\Delta H^{\bm k}_{\rm{\mr{FE}}}=
-B_{\mr{eff}}({\bm k})\,\,\frac{1}{2}
C_{\bm k}^\dagger \sigma_z C_{\bm k}
+\order{\omega^{-4}},
\label{eq:DeltaFH}
\end{align}
From Eqs~(\ref{SzCW}) and (\ref{eq:DeltaFH}), 
we find the relation $\langle\langle s^z_{\bm k}\rangle\rangle\propto \frac{\alpha_{\rm R}^2E_0^2}{\omega^3}$, which leads to
\begin{align}
    \langle\langle s^z\rangle\rangle=&
    \frac{1}{N}\langle\langle S^z_{\rm tot}\rangle\rangle
    \propto \frac{\alpha_{\rm R}^2E_0^2}{\omega^3}
    \label{eq:mag_power}
\end{align}
at least in a sufficiently 
high-frequency regime. This power-law behavior has already been predicted by the theories of IFE in isolated electron systems~\cite{pershan1966,tanaka2020}. 
Equation~(\ref{eq:mag_power}) shows that the same power law for the laser-driven magnetization holds even in laser-driven ``dissipative'' electron systems.

We here note that the power law of $ \langle\langle s^z\rangle\rangle\propto E_0^2/\omega^3$ holds for the off-resonant IFE in standard metallic systems considered in the present study, while other laser-field and frequency dependencies of the laser-driven magnetization can occur, depending on the sorts of systems such as a resonant-type IFE~\cite{battiato2014,berritta2016}, Mott insulators~\cite{banerjee2022} and magnetic insulators~\cite{takayoshi2014,takayoshi2014a,ikeda2020,ikeda2021}.

\begin{figure}[b]
    \centering
    \includegraphics[keepaspectratio, width=\linewidth]{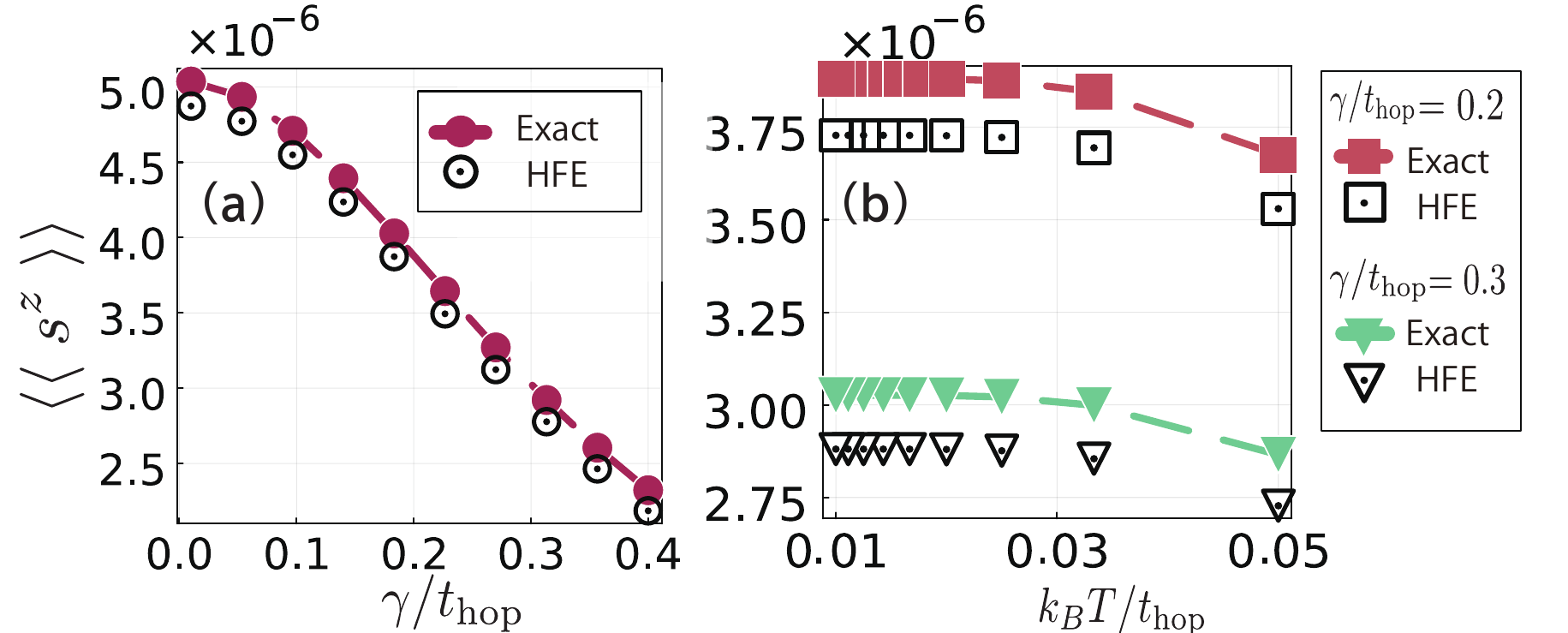}
  \caption{Laser-driven magnetization $\langle \ev{s^z}\rangle$ as a function of (a) the dissipation magnitude ($\gamma$) and (b) temperature ($T$). Other parameters are chosen to be $eaE_0/t_{\mr{hop}}=0.1$, $\alpha_{\rm R}/t_{\mr{hop}}=0.1$, and $\hbar\omega /t_{\mr{hop}}=1.0$. In (a), $\gamma$ changes in the range of 
  $0<\gamma/t_{\rm hop}\leq 0.4$ at $T=0$. 
  Red points and white circles, respectively, correspond to the result of the numerically solved GKSL equation and that of the Floquet high-frequency expansion [see Eqs.~(\ref{SzCW}) and (\ref{eq:mag_power})]. In panel (b), we depict the results of $\gamma/t_{\rm hop}=0.2$ and 0.3. Red and green points are numerical results of the GKSL equation, while white squares and triangles are those of the high-frequency expansion. Temperature is changed from $k_BT/t_{\mr{hop}}=0.0$ to $0.05$, which corresponds to about \SI{500}{K} for $t_{\rm hop}=$ \SI{1}{eV}.
  }
\label{fig:T_gam}
\end{figure}

\subsection{\label{sec:Ness-GamTem} \texorpdfstring{$\gamma$}{gamma} and \texorpdfstring{$T$}{T} Dependence}
In the remaining part of Sec.~\ref{sec:Ness}, we will discuss the properties of the NESS with numerical computation of the GKSL equation. 
This subsection is devoted to the dissipation ($\gamma$) and temperature ($T$) dependence of the laser-driven magnetization $\langle\langle s^z\rangle\rangle$. We stress that the GKSL formalism makes it possible to discuss the $\gamma$ and $T$ dependence whereas a standard method based on Schr\"odinger equation cannot treat the effects of dissipation and temperature. 

Figure~\ref{fig:T_gam}(a) shows the $\gamma$ dependence of the numerically computed spin moment per one site $\langle\langle s^z\rangle\rangle$ at $T=0$. On the other hand, 
$\langle\langle s^z\rangle\rangle$ 
can also be computed as Eqs.~(\ref{SzCW}) and (\ref{eq:mag_power}) under the condition of a sufficiently high frequency. 
In Fig.~\ref{fig:T_gam}, we plot this analytic result of the Floquet theory as well. 
One sees that (as expected) the value of $\langle\langle s^z\rangle\rangle$ monotonically decreases with the growth of the dissipation strength $\gamma$ and the analytic result well agrees with the accurate numerical one. 
It is also shown that even if $\gamma$ becomes 
close to the order of $t_{\rm hop}$, the laser-driven magnetization $\langle\langle s^z\rangle\rangle$ still remains 
at the same order as the value at $\gamma\to 0$. 
We have verified that the magnetization at the limit of $\gamma\to 0$ and $T=0$ is in agreement with that in the previous study~\cite{tanaka2020}, which is computed from the solution of Schr\"odinger equation.

Next, we consider the $T$ dependence of $\langle\langle s^z\rangle\rangle$. 
As we mentioned in Sec.~\ref{sec:Mode-GKSL}, 
in the case of $T=0$, it is enough to analyze the GKSL equations in the doughnut regime between two Fermi surfaces shown in Fig.~\ref{fig:HamModel_cw}(d). At finite temperatures, the possibilities of the appearance of one-particle states becomes finite in full Brillouin zone. However, if we consider a sufficiently low temperature regime ($k_{\rm B}T\ll t_{\rm hop}$), the effect of the area outside the doughnut regime would be still negligible for the analysis of $\langle\langle s^z\rangle\rangle$. 
Under this simple approximation, we here discuss the $T$ dependence of $\langle\langle s^z\rangle\rangle$ by solving the GKSL equations only in the doughnut regime. 
Figure~\ref{fig:T_gam}(b) represents the $T$ dependence of numerically computed $\langle\langle s^z\rangle\rangle$ and the same quantity computed by the Floquet high-frequency expansion. One sees that these numerical and analytical results agree with each other in a semi-quantitative level. 
The laser-driven spin moment is shown to monotonically decrease with increasing $T$, while the figure also shows that the laser-driven magnetization is stable against temperature change if $T$ is small enough compared with $t_{\rm hop}$, i.e., the typical energy scale of the electron system. 
It is found that even when $T$ is chosen to be room temperature, the laser-driven magnetization takes the value of the same order as at $T=0$ for $t_{\rm hop}=1$eV.

Finally, we shortly remark the $\gamma\to 0$ limit, i.e, the isolated system. From Fig.~\ref{fig:T_gam}(a), we see that the NESS at $\gamma\to 0$ seems to smoothly connect to the NESS with a finite $\gamma$. However, as we mentioned in the Introduction, the theoretical analyses in Refs.~\cite{lazarides2014,dalessio2014} show that if we apply a laser to an isolated many-body system for a long time, the system generally approaches to a featureless, high-temperature-like state. The emergence of a NESS at $\gamma\to 0$ is inconsistent with this statement in Refs.~\cite{lazarides2014,dalessio2014}. The NESS realization at $\gamma\to 0$ might be an artifact owing to the use of a free-fermion model. In real experiments of laser application, however, even ideal metals are expected to reach a featureless state because of the effects of a weak but finite interaction, impurities, crystalline defects, etc.

\subsection{\label{sec:Ness-AC} AC-field Dependence}
\begin{figure}[t]
    \centering
    \includegraphics[keepaspectratio, width=\linewidth]{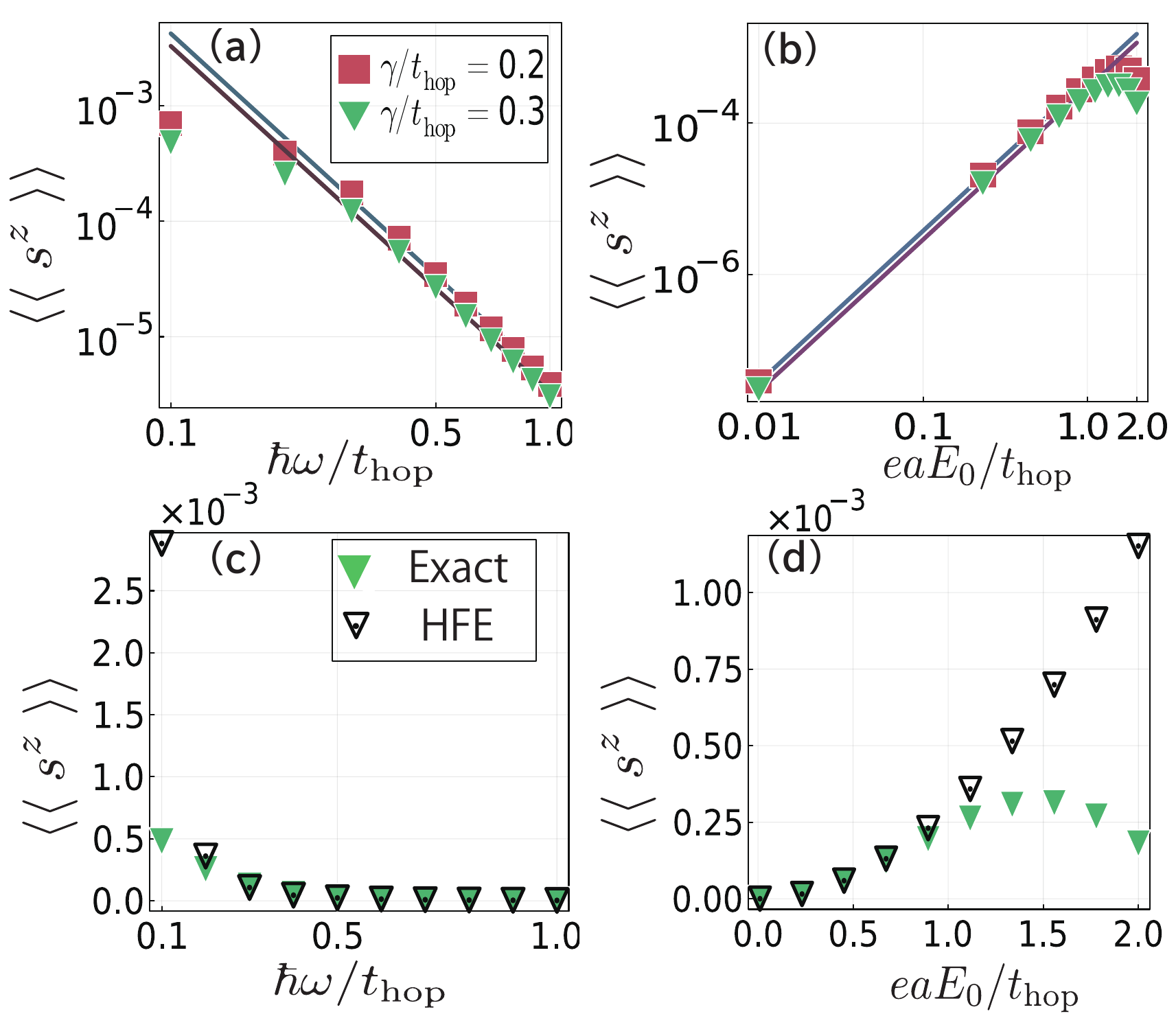}
    \caption{[(a), (b)] Log-log plot of $\langle \ev{s^z} \rangle$ as a function of laser frequency $\omega$ (a) and intensity $E_0$ (b). Red and green points respectively represent the results of numerical calculations for $\gamma/t_{\mr{hop}}=0.2$ and $0.3$. The solid lines are fitting curves obeying the power law $E_0^2/\omega^3$. Other parameters are set to $\alpha_{\rm R}/t_{\mr{hop}}=0.1$, $k_BT/t_{\mr{hop}}=0$ and $eaE_0/t_{\mr{hop}}=0.1$. For instance, $eaE_0/t_{\mr{hop}}=0.1$ corresponds to $E_0=$\SI{2}{MVcm^{-1}} and $\hbar\omega /t_{\mr{hop}}=1.0$ does to $\omega/2\pi=0.24$PHz when $t_{\mr{hop}}=$\SI{1}{eV}. 
    [(c), (d)] Linear plot of $\langle \ev{s^z} \rangle$ as a function of $\omega$ (c) and $E_0$ (d). Panel (c) [(d)] corresponds to panel (a) [(b)]. 
    Green triangles represent the numerical calculation results for $\gamma/t_{\rm hop}=0.3$ and white triangles the results of the Floquet high-frequency expansion. 
    }
    \label{Fig:CW_omegaE0}
\end{figure}
In this subsection, we consider the effects of the frequency $\omega$ and the intensity $E_0$ of the AC electric field on the laser-driven magnetization in the NESS. 
Figure~\ref{Fig:CW_omegaE0} shows the $\omega$ and $E_0$ dependence of the numerically computed $\langle\langle s^z\rangle\rangle$ at $T=0$. 
As we already have shown in Sec.~\ref{sec:Ness-Floquet}, 
$\langle\langle s^z\rangle\rangle$ is proportional to $E_0^2/\omega^3$ for $|E_0/(\hbar\omega)|\ll 1$, i.e., for a sufficiently high-frequency regime [see Eq.~(\ref{eq:mag_power})]. 
Figures~\ref{Fig:CW_omegaE0}(a) and \ref{Fig:CW_omegaE0}(b) clearly indicate the power law holds in the high-frequency regime, while $\langle\langle s^z\rangle\rangle$ deviates from the law when $|E_0/(\hbar\omega)|$ becomes larger. In fact, one finds from Figs.~\ref{Fig:CW_omegaE0}(c) and \ref{Fig:CW_omegaE0}(d) that the numerically computed exact value of $\langle\langle s^z\rangle\rangle$ deviates from the result of the Floquet high-frequency expansion for $|E_0/(\hbar\omega)|\agt 1$. 
As we already mentioned, this power law of the laser-driven magnetization is the same as that in dissipationless IFE~\cite{lp1961,pershan1966,tanaka2020}.

In experiments of IFE, ultraviolet to infrared laser has been usually used. Roughly speaking, their photon energy is the same order as the energy scale of solid electron systems, i.e., $\hbar\omega\sim t_{\rm hop}$. Figure~\ref{Fig:CW_omegaE0} tells us that in the case of $\hbar\omega\sim t_{\rm hop}$, the magnitude of the AC electric field should increase up to $eaE_0\sim t_{\rm hop}$
to maximize $\langle\langle s^z\rangle\rangle$. 
However, we have to note that in experiments, $E_0$ can approach at most 1 to 10 MV/cm, in which $eaE_0$ is usually much smaller than $t_{\rm hop}$.

\section{\label{sec:Pulse} Laser Pulses}
In the previous section, we have studied properties of the NESS that occurs after a long-time application of CW laser. The NESS is very useful to capture the essential, universal features of FE phenomena including IFE. However, short laser pulses (not CW) have been widely used in experiments of IFE~\cite{vanderziel1965, kimel2005a,hansteen2006, satoh2010,makino2012}. Therefore, here we explore the ultrafast spin dynamics driven by a short pulse of circularly polarized laser. Here, ``ultrafast'' spin dynamics means that it is faster than typical time scale of electronics, while (as one will see soon later) it is slower than the laser period $T_\omega$. 

The electric field of the laser pulse was already defined in Eq.~(\ref{Eq:Pulse}). The pulse length $\tau$ and the initial time $t_{\rm ini}$ are respectively fixed to {about $14\hbar/t_{\mr{hop}}$} and $-600\hbar/t_{\mr{hop}}$ in our numerical calculations. We explore the pulse induced IFE by numerically solving the GKSL equation for laser-pulse driven electron systems.
For simplicity, we focus on the zero temperature case ($T=0$) in this section.

\subsection{\label{sec:FerroMetal}Ferromagnetic Metal}
\begin{figure}[b]
\centering\includegraphics[keepaspectratio, width=\linewidth]{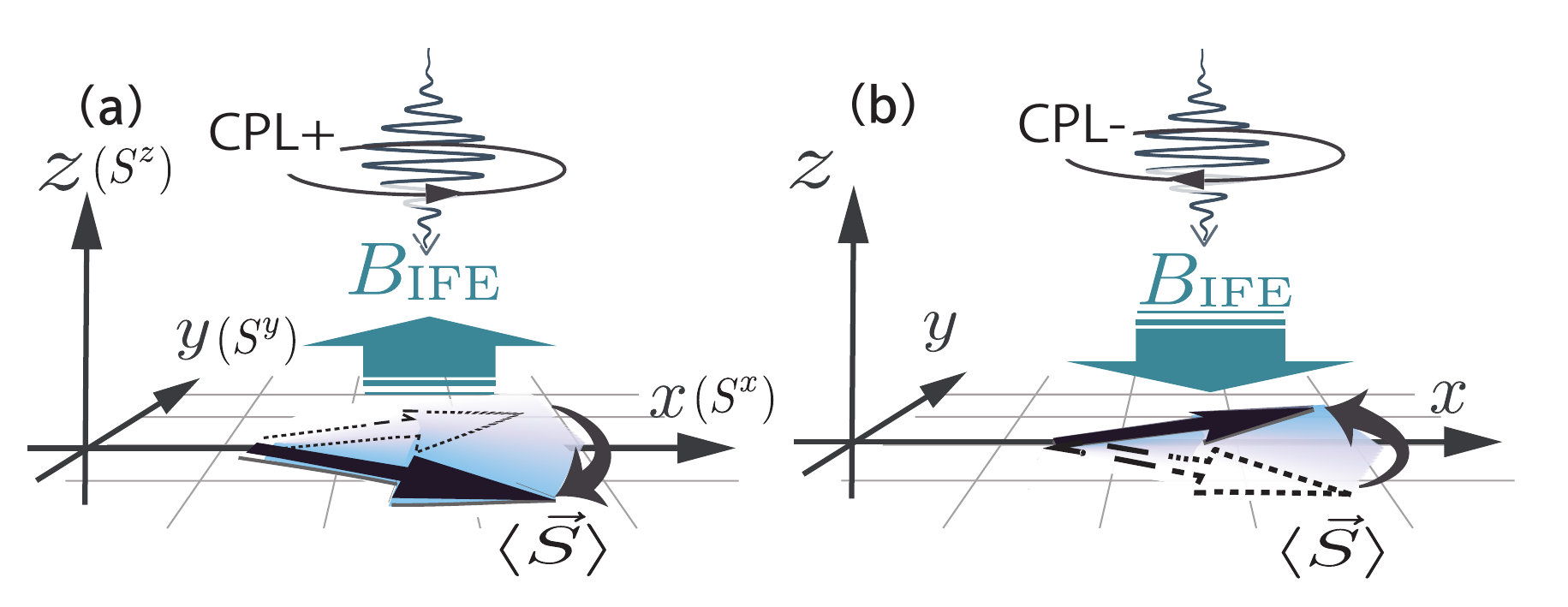}
  \caption{
Schematic image of IFE driven by pulse of circularly polarized laser (CPL) with (a) right- or (b) left handedness. The laser-driven 2D electron system on the $x$-$y$ plane has a finite ferromagnetic moment along the $S^x$ axis and the incident pulse enters along the $z$ axis. The arrow of $B_{\mr{IFE}}$ means the direction of an instantaneous magnetic field induced by the pulse IFE. The direction of $B_{\mr{IFE}}$ is expected to be controlled by changing the light polarization from right (left) to left (right) circular types. Black arrows stand for the expected precession motion of the spin moment. 
}
\label{Fig:PulsePonchi}
\end{figure}
In the case of CW laser, electron spins are polarized along the effective magnetic field created by the circularly polarized laser, as shown in Eqs.~(\ref{SzCW})-(\ref{eq:mag_power}). However, when a short laser pulse is applied to electron systems, a pulse-driven magnetic field immediately disappears before the electron spins becomes polarized. Instead of spin polarization, a precession of the magnetic moment is induced by the pulse-driven instantaneous magnetic field if we consider IFE in a metallic system with a magnetic order. Such a precession has been observed in several experiments of IFE~\cite{kimel2005a,satoh2010,makino2012}. 
Let us consider the setup of IFE shown in Fig.~\ref{Fig:PulsePonchi}: a laser pulse with circular polarization is applied to a 2D ferromagnetic metal with a ferromagnetic moment along the $x$ axis. The laser-driven instantaneous magnetic field ($B_{\rm IFE}$) is parallel to the $z$ axis and its direction becomes positive and negative, depending on the helicity (right- and left handedness) of laser. 
To theoretically discuss the ultrafast precession, we hence should prepare a magnetically ordered electron state. For simplicity, we focus on a ferromagnetic metal state like Fig.~\ref{Fig:PulsePonchi}. 
To this end, we extend the 2D paramagnetic Rashba model of Eq.~(\ref{eq:hamCW}) to a 2D electron model with a ferromagnetic moment, whose Hamiltonian is defined as   
\begin{equation}
\label{Eq:hamPulse}
    {H}_{\mr{FM}}=H_{\rm{PM}}
    +H_{\mr{MFT}}, 
\end{equation}
where $H_{\mr{MFT}}$ is given by
\begin{align}
\label{MFT}
    H_{\mr{MFT}}=-B_x S_{\rm tot}^x=-B_x\sum_{\bm{k}}\frac{1}{2}C_{\bm{k}}^\dagger\sigma_x C_{\bm{k}}. 
\end{align}
The index ``FM'' means ``ferromagnetic'' and 
the final term $H_{\mr{MFT}}$ is an Zeeman coupling due to a magnetic field $B_x$ and has been introduced to generate a finite ferromagnetic moment along the $S^x$ axis like Fig.~\ref{Fig:PulsePonchi}. 
One may consider that this Zeeman term emerges from a mean-field treatment for electron-electron interactions including Coulomb interaction, Hund coupling, etc \cite{white2007}. 
However, we here regard the effective field $B_x$ as a merely free parameter to realize a ferromagnetic metal state.  
In this section, we use this mean-field Hamiltonian to investigate the ultrafast spin dynamics driven by laser pulses.

The model of Eq.~(\ref{Eq:hamPulse}) can be easily solved because it is a free-fermion type. 
In the $\bm k$ space, the Hamiltonian reads
\begin{align}              
H_{\rm FM}=\sum_{\bm{k}}
C^\dagger_{\bm k}
    \mqty(
    \varepsilon_{\bm{k}} & \eta_{\bm k}-B_x/2\\
    \eta_{\bm k}^*-B_x/2 & \varepsilon_{\bm{k}} )
C_{\bm k}.
\label{eq:ferro_k}
\end{align}
From this Hamiltonian, one can compute expectation values of arbitrary observable in the equilibrium state. 

\begin{figure}[t]
\centering\includegraphics[keepaspectratio, width=\linewidth]{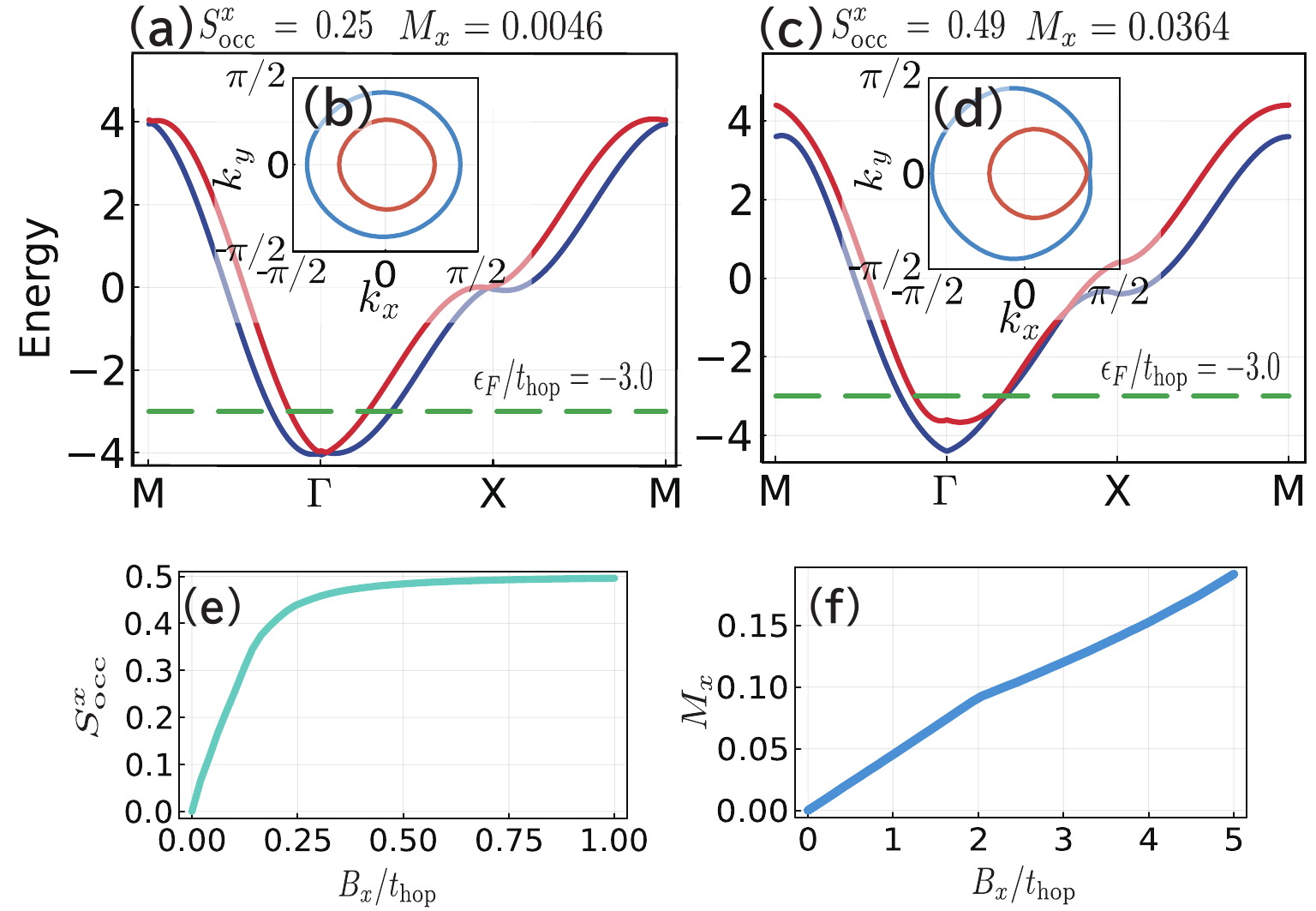}
  \caption{[(a), (c)] Energy bands of the mean-field ferromagnetic model of Eq.~(\ref{Eq:hamPulse}) at (a) a weak spin moment $M_x=0.08$ and (c) an almost saturated moment $M_x=0.49$. Parameters are set to $t_{\rm hop}=$\SI{1}{eV} and $\alpha_{\rm R}/t_{\mr{hop}}=0.5$. The Fermi energy is fixed to $\epsilon_F=-3t_{\rm hop}$. 
  [(b),(d)] Panels (b) and (d) respectively correspond to the Fermi surfaces for the cases (a) and (c). [(e), (f)] Magnetization curves of (e) $S^z_{\rm occ}$ and (f) $M_x$ as a function of $B_x$ for $\alpha_{\rm R}/t_{\mr{hop}}=0.5$ and $k_BT/t_{\mr{hop}}=0.0$. 
}
\label{fig:HamModel_Pulse}
\end{figure}

As the magnetization induced by $B_x$ is important in this section, we here introduce two expectation values associated to the spin moment:
\begin{align}
    M_\alpha=&\langle s^\alpha\rangle_{t_{\rm ini}},\nonumber\\
    S^\alpha_{\rm occ}=&\frac{1}{N_{\rm occ}}\sum_{{\bm k}\in{\rm occ}}\langle s_{\bm k}^\alpha\rangle_{t_{\rm ini}}.
    \label{eq:ferromagnetization}
\end{align}
Here, $M_\alpha$ stands for the $\alpha$ component of magnetization per site at the initial time $t=t_{\rm ini}$, i.e., the magnetization in an equilibrium state before the application of a laser pulse. On the other hand, $N_{\rm occ}$ is total number of the occupied one-particle state at $T=0$ and $\sum_{{\bm k}\in{\rm occ}}$ means the summation over all the one-particle states at $T=0$, which is equivalent to the doughnut area between two Fermi surfaces of the model $H_{\rm FM}$. 
Therefore, $S^\alpha_{\rm occ}$ indicates how many electron spins are polarized along the $\alpha$ axis in all the one-particle states in the Brillouin zone: The saturation value $S^\alpha_{\rm occ}=1/2$ corresponds to the state where all the electron spins in one-electron occupied states are fully polarized at $T=0$. 
Figures~\ref{fig:HamModel_Pulse}(a)-\ref{fig:HamModel_Pulse}(d) show the energy bands and Fermi surfaces of the ferromagnetic metal model of Eq.~(\ref{eq:ferro_k}) with finite mean fields $B_x$ at Fermi energy $\epsilon_F=-3t_{\rm hop}$. 
One sees that Fermi surfaces gradually change with $B_x$ increased. Figure~\ref{fig:HamModel_Pulse}(e) and \ref{fig:HamModel_Pulse}(f) are respectively the magnetization curves of $S_{\rm occ}^x$ and $M_x$ as a function of $B_x$ at $T=0$. 
They tell us that $S_{\rm occ}^x$ is almost saturated for $B_x>0.5t_{\rm hop}$, while $M_x$ still monotonically increase together with $B_x$ even if $B_x$ is beyond $5t_{\rm top}$. When $B_x$ is much larger than $t_{\rm hop}$ ($B_x\gg t_{\rm hop}$), two energy bands $E_1^{\bm k}$ and $E_2^{\bm k}$ are massively separated. As a result, the lower band is completely occupied by electrons with $S^x=+1/2$ polarization ($E_1^{\bm k}<\epsilon_F$) and the higher band is empty ($E_2^{\bm k}>\epsilon_F$). 
This situation corresponds to the saturation of $M_x$. 
However, we want to consider a realistic ferromagnetic metal state within our simple mean-field model.
In real ferromagnetic metals~\cite{white2007}, the deviation between spin-$\uparrow$ and spin-$\downarrow$ electron numbers is usually relevant only near the Fermi surface. 
From this argument and Fig.~\ref{fig:HamModel_Pulse}(e),
we should tune the value of the mean field, for example, in a range $0.01<B_x/t_{\rm hop}<0.2$ such that $S_{\rm occ}^x$ takes a moderate value far from the saturation value $1/2$. 
In fact, as we will explain in Appendix~\ref{App:numerical}, if we start from almost saturated ferromagnetic state with $S_{\rm occ}^x\simeq 1/2$, 
even strong laser pulse can induce a quite small precession motion of the spin moment because the spin moment is strongly locked due to the considerably large field $B_x$. 

When the effect of laser pulse is introduced in Eq.~(\ref{Eq:hamPulse}), it is enough to replace $H_{\rm PM}$ to $H_{\rm PM}(t)$ using the Peierls formalism like Sec.~\ref{sec:Mode-Laser}. However, we should note that we use the vector potential for the laser ``pulse'' field of Eq.~(\ref{Eq:Pulse}) instead of CW laser. 
The time-dependent pulse-driven Hamiltonian is given by 
\begin{equation}
\label{Eq:timedepH_pulse}
    {H}_{\mr{FM}}(t)=H_{\rm{PM}}(t)
    +H_{\mr{MFT}}. 
\end{equation}
This Hamiltonian is still $\bm k$-diagonal like the case of CW laser. Using Eq.~(\ref{Eq:timedepH_pulse}), we will numerically solve the GKSL equation. 

\subsection{\label{sec:Pulse-Prece} Pulse induced Precession}
\begin{figure*}[t]
    \includegraphics[width=0.9\textwidth]{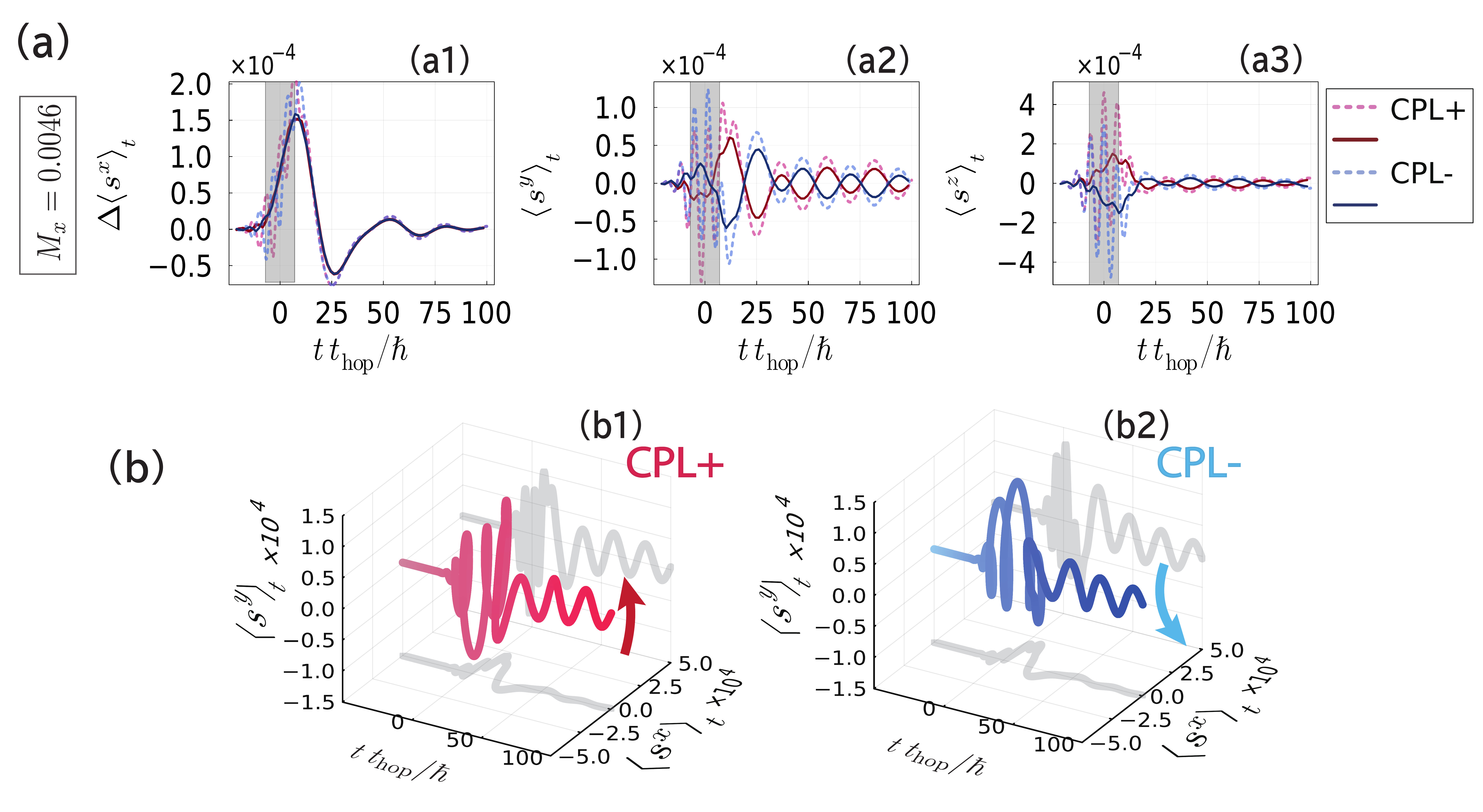}    
  \caption{[(a1)-(a3)] Time evolution of $\ev {s^x}_t$, $\ev {s^y}_t$, and $\ev {s^z}_t$ computed by the GKSL equation for a ferromagnetic metal state in the model~(\ref{Eq:timedepH_pulse}) with ``magnetization'' $S_{\rm occ}^x\simeq 0.248$ (corresponds to $M_x\simeq 0.0046$ and $B_x/t_{\rm hop}=0.1$) at $T=0$. Here, $\Delta\ev {s^\alpha}_t=\ev {s^\alpha}_t-\ev {s^\alpha}_{t_{\rm ini}}$ and $\ev {s^{y,z}}_{t_{\rm ini}}=0$. 
  Dotted lines denote $\ev {s^\alpha}_t$ themselves, while solid lines are the slow modes defined by $s_{\rm slow}^\alpha(t)=\frac{1}{2T_\omega}\int_{t-T_\omega}^{t+T_\omega} dt' \Delta\langle s^\alpha\rangle_{t'}$. Red and blue colors respectively correspond to right circularly polarized ($\hbar\omega/t_{\mr{hop}}=1$) and left circularly polarized ($\hbar\omega/t_{\mr{hop}}=-1$) pulses. The gray region denotes the width of laser pulse $\tau$: The laser intensity is strong enough in this range. 
  [(b1), (b2)] Plot of slow mode of (a1)-(a3) in the three-dimensional space $(t t_{\mathrm{hop}}/\hbar
  , \ev {s^x}_t, \ev {s^y}_t)$. 
  Other parameters are chosen to be  $eaE_0=0.5$, $\alpha_{\rm R}/t_{\mr{hop}}=0.1$, $k_BT/t_{\mr{hop}}=0$, $\gamma/t_{\mr{hop}}=0.01$, and $\tau\simeq 14\hbar/t_{\mr{hop}}$. 
}
\label{PulseMxLowhigh}
\end{figure*} 
In this subsection, we discuss the numerical results of the spin moment induced by pulse laser. 
For simplicity, the numerical computation will all be done at $T=0$ in this section. 
Figure~\ref{PulseMxLowhigh}(a) shows the spin moment of $\langle s^\alpha\rangle_t$ as a function of time $t$ in a ferromagnetic metal state with a moderate value of $S^x_{\rm occ}$.
The dotted lines represent the actual time evolution, while the solid lines represent the extracted slow modes that is defined as
\begin{align}
    s_{\rm slow}^\alpha(t)=\frac{1}{2T_\omega}\int_{t-T_\omega}^{t+T_\omega} dt' \Delta\langle s^\alpha\rangle_{t'},
\end{align}
where we have taken a time average over $2T_\omega$ and we have introduced $\Delta\ev {s^\alpha}_t=\ev {s^\alpha}_t-\ev {s^\alpha}_{t_{\rm ini}}$ and $\ev {s^{y,z}}_{t_{\rm ini}}=0$. 
The gray area indicates the full width at half maximum of the laser pulse, namely, the time interval during which laser intensity is strong enough.

The most remarkable part in Fig.~\ref{PulseMxLowhigh}(a) is the behavior of the slow mode of $\langle s^y\rangle_t$. From Fig.~\ref{Fig:PulsePonchi}, we can expect that a clear precession arises in the $y$ component of spin in our setup and the initial ``phase'' of the precession driven by a right-handed light ($\omega>0$) deviates by $\pi$ from that by a left-handed one ($\omega<0$). 
One finds this phase difference in $\langle s^y\rangle_t$ of Fig.~\ref{PulseMxLowhigh}(a2). 
The phase difference has been indeed observed in several experiments of pulse-driven IFE~\cite{ kimel2005a,satoh2010,makino2012} and is a definite evidence for the emergence of an instantaneous magnetic field ($B_{\rm IFE}$ in Fig.~\ref{Fig:PulsePonchi}) by a circularly polarized laser pulse. 
In real magnetic materials, the characteristic frequency of the precession (i.e., slow mode) is determined by the spin-wave eigenenergy~\cite{white2007}. 
Our model does not include the correlation between spin moments in neighboring sites and hence cannot reproduce the precession with the spin-wave frequency. To take the spin-wave nature into account in the microscopic level, we have to analyze laser-driven dynamics in correlated electron systems on a lattice such as Hubbard models, more realistic multi-band correlated electron models, etc. 
It is an important future issue of the research of IFE. We however emphasize that a slow precession and the phase difference can be captured within our free-fermion model for a ferromagnetic metal. Here, we again note that the slow precession is fast compared with typical time scale of electronics. 
``Slow'' means that it is slower than the laser frequency and we may refer to this mode as ``ultrafast'' spin dynamics. 

Figure~\ref{PulseMxLowhigh}(a) also demonstrates that 
the slow modes of $\langle s^x\rangle_t$ for right- and left-handed lights are almost degenerate. 
This behavior can be understood from Fig.~\ref{Fig:PulsePonchi}. Namely, if the time evolution of $\langle s^\alpha\rangle_t$ is sufficiently close to the ideal precession 
on the $S^x$-$S^y$ plane like Fig.~\ref{Fig:PulsePonchi}, a phase difference does not appear in $\langle s^x\rangle_t$ and 
it arises only in $\langle s^y\rangle_t$. 
One can also find from Fig.~\ref{PulseMxLowhigh}(a3) that $\langle s^z\rangle_t$ slightly increases (decreases) during the application of right-handed (left-handed) laser pulse. 
This could be interpreted as the growth of spin moment by the instantaneous magnetic field $B_{\rm IFE}$. 

Figure~\ref{PulseMxLowhigh}(b) depicts the trajectories of the pulse-driven slow dynamics in the three-dimensional $(t t_{\rm hop}/\hbar, \ev {s^x}_t, \ev {s^y}_t)$ space, from which one can visually understand the precessions generated by right and left circularly polarized pulses. The corresponding movie is in Ref.~\footnote{See Movie\_Supp.mp4}.

\begin{figure*}[t]
    \centering
    \includegraphics[width=0.95\textwidth]{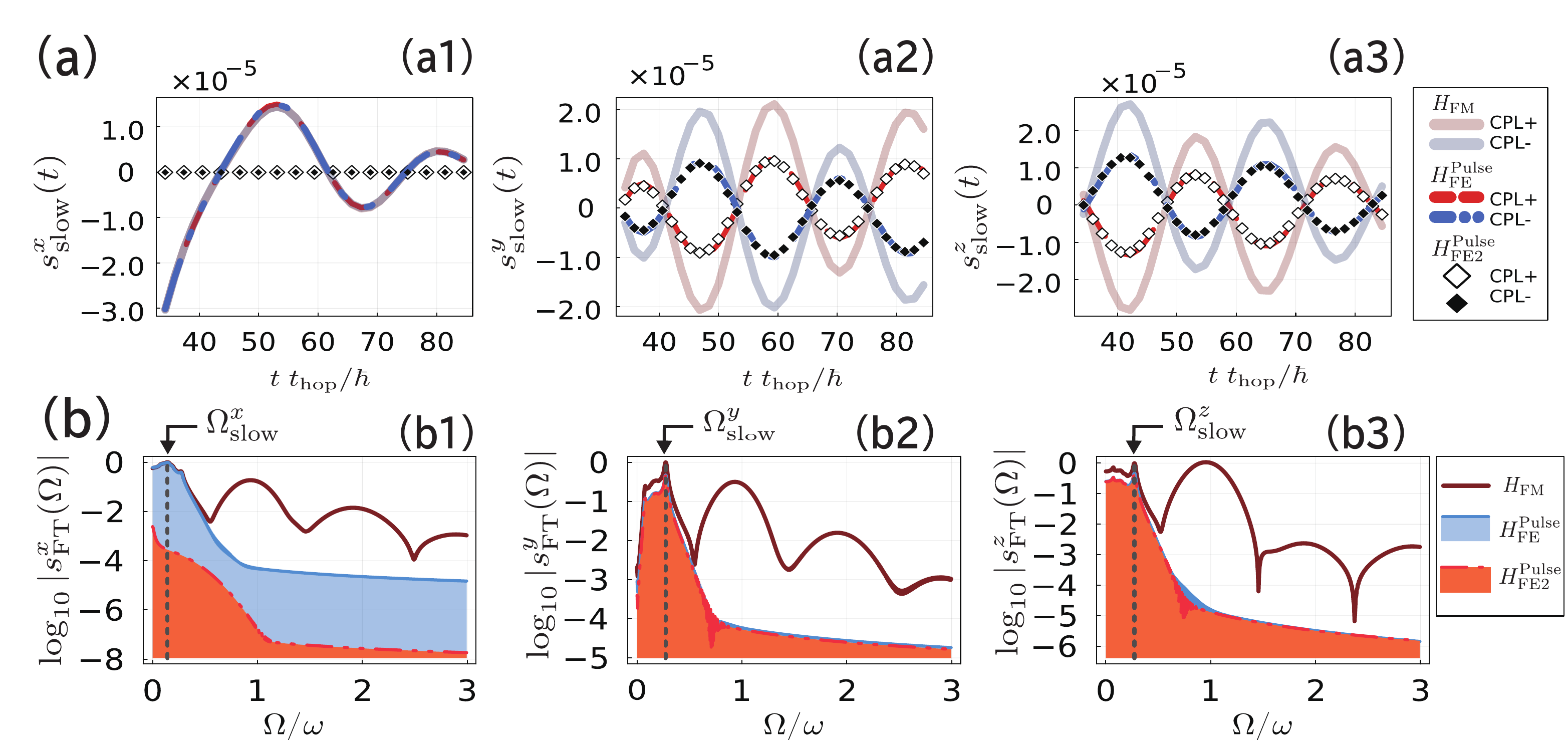}
  \caption{
[(a1)-(a3)] 
Time evolution of slow modes of $\ev {s^x}_t$, $\ev {s^y}_t$, and $\ev {s^z}_t$ in the case of circularly polarized laser pulse at $T=0$. 
Each panel includes the numerical results estimated by three models of Eqs.~(\ref{Eq:timedepH_pulse}),  (\ref{hhefPulse}), and (\ref{hhefPulse2}). 
CPL$+$ and CPL$-$ respectively stand for right ($\hbar\omega/t_{\mr{hop}}=1>0$) and left ($\hbar\omega/t_{\mr{hop}}=-1<0$) circularly polarized laser pulses. 
Light red (CPL$+$) and light blue (CPL$-$) curves are the results of Eq.~(\ref{Eq:timedepH_pulse}). Orange dotted (CPL$+$) and blue dotted (CPL$-$) lines are those of Eq.~(\ref{hhefPulse}). 
White diamond (CPL$+$) and black diamond (CPL$+$) marks are those of Eq.~(\ref{hhefPulse2}). 
[(b1)-(b3)] Fourier spectra of 
$s^\alpha_{\mr{FT}}$
as a function of arbitrary frequency $\Omega$ in the case of right circularly polarized pulse. 
We use the log plot and the value of the vertical axis is 
measured from the maximum value of $\log_{10}|s^\alpha_{\mr{FT}}(\Omega)|$. Red line is the result of Eq.~(\ref{Eq:timedepH_pulse}), blue area is that of Eq.~(\ref{hhefPulse}), and orange area is that of Eq.~(\ref{hhefPulse2}). 
In (b1)-(b3), we define the characteristic angular frequency $\Omega_{\mr{slow}}^\alpha$ that is the highest peak position of $\log_{10}|s^\alpha_{\mr{FT}}(\Omega)|$ in the low-frequency regime $\Omega<\omega$. Other parameters are chosen to be $M_x\simeq 0.0046$ ($S_{\mr{occ}}^x\simeq 0.25$ and $B_x/t_{\mr{hop}}=0.1$), $t_{\rm top}=1$,
$eaE_0/t_{\mr{hop}}=0.5$, 
$\alpha_{\rm R}/t_{\mr{hop}}=0.1$, and  
$\gamma/t_{\mr{hop}}=0.01$.
  }
\label{Fig:Models_time}
\end{figure*}
\subsection{\label{sec:Pulse-Timedepen} Time-dependent Effective Hamiltonian}
Here, we consider how one can understand the  precession mode of Fig.~\ref{PulseMxLowhigh} from the Floquet-theory perspective. 
For this purpose, let us first remember the case of a CW laser. In the case, the Floquet high-frequency expansion~\cite{eckardt2015,mikami2016} enables us to lead to the time-independent Floquet Hamiltonian, which is given by 
\begin{multline}
\label{Eq:hhEFCW}
H_{\mr{FE}}^{\mr{CW}}=\qty(1-\frac{E_0^2}{4\omega^2})H_{\rm{PM}}\\
-\sum_{\bm k}B_{\mr{eff}}(\bm k)\frac{1}{2}C_{\bm k}^\dagger \sigma_z C_{\bm k}+H_{\mr{MFT}},
\end{multline}
where $B_{\rm eff}=2\alpha_{\rm R}^2E_0^2/\omega^3\cos(k_x)\cos(k_y)$ [see Eq.~(\ref{eq:FH})]. 
In the case of laser pulse, on the other hand, the high-frequency expansion is no longer applicable in principle. 
However, we can expect that the expansion is still valid for a short time interval, which is somewhat longer than the laser period $T_\omega$. 
Under this rough expectation, we may introduce the effective Hamiltonian (time-evolution operator) for the case of laser pulse, by replacing $E_0$ with $E_0 e^{-t^2/(\tau/2)^2}$ in Eq.~(\ref{Eq:hhEFCW}). 
Therefore, the effective Hamiltonian for laser pulse is given by
\begin{multline}
\label{hhefPulse}
H_{\mr{FE}}^{\mr{Pulse}}(t)=
\qty(1-\frac{(e^{-2t^2/\tau^2}E_0)^2}{4\omega^2})H_{\rm{PM}}\\
-e^{-4t^2/\tau^2}
\sum_{\bm k}B_{\mr{eff}}(\bm k)\frac{1}{2}
C_{\bm k}^\dagger \sigma_z C_{\bm k}+H_{\mr{MFT}}.
\end{multline}
Furthermore, the laser-driven magnetic field $B_{\rm eff}(\bm k)$ is expected to be more significant rather than the correction to $H_{\rm PM}$. Hence, we also introduce another effective Hamiltonian: 
\begin{align}
\label{hhefPulse2}
H_{\mr{FE}2}^{\mr{Pulse}}(t)=&
H_{\rm{PM}}-e^{-4t^2/\tau^2}
\sum_{\bm k}B_{\mr{eff}}(\bm k)\frac{1}{2}
C_{\bm k}^\dagger \sigma_z C_{\bm k}
\nonumber\\
&+H_{\mr{MFT}}.
\end{align}
These two Hamiltonians, $H_{\mr{FE}}^{\mr{Pulse}}(t)$ and $H_{\mr{FE}2}^{\mr{Pulse}}(t)$, are expected to describe the slow dynamics, whose time scale is slower than the laser period $T_\omega$.

We compare the numerically exact results of $\langle s^\alpha\rangle_t$ with those derived from $H_{\mr FE}^{\rm Pulse}(t)$ or $H_{\mr{FE}2}^{\mr{Pulse}}(t)$. To this end, we here define the Fourier transform of the magnetization as  
\begin{align}
s_{\mr{FT}}^\alpha(\Omega)=\frac{t_{\rm hop}}{\hbar}\int^{\infty}_{-\infty}\Delta\ev{s^\alpha}_t e^{-i\Omega t}dt,
\end{align}
where $\Delta\ev {s^\alpha}_t=\ev {s^\alpha}_t-\ev {s^\alpha}_{t_{\rm ini}}$ and $\ev {s^{y,z}}_{t_{\rm ini}}=0$. The factor $\frac{t_{\rm hop}}{\hbar}$ is introduced to make $s_{\mr{FT}}^\alpha(\Omega)$ dimensionless. 
\begin{figure}[t]
    \centering
    \centering
    \includegraphics[width=0.5\textwidth]{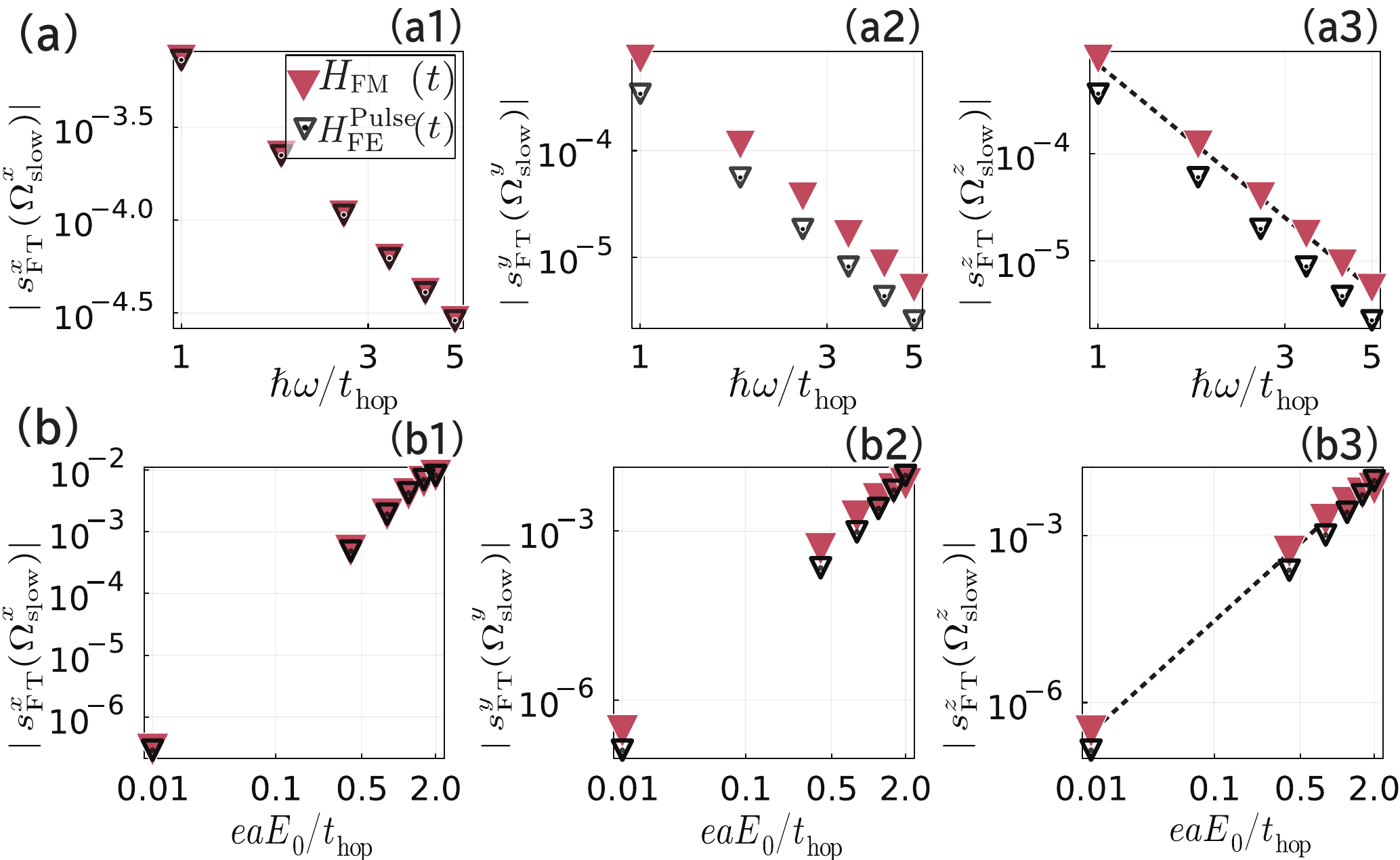}
    \caption{[(a),(b)] Log plot of $|s_{\mr{FT}}^\alpha(\Omega_{\mr{slow}}^\alpha)|$ as a function of the frequency $\omega$ (a) and the field strength $E_0$ (b) of laser pulse. The frequency $\Omega_{\mr{slow}}^{\alpha}$ is determined as the highest peak position of $|s_{\mr{FT}}^\alpha(\Omega)|$ in $\Omega<\omega$. Red [white] triangles are the numerical result of combining the original mean-field Hamiltonian $H_{\rm FM}(t)$ [the effective Hamiltonian $H_{\rm FE}^{\rm Pulse}(t)$] with the GKSL equation. 
    In panels (a1)-(a3), the laser intensity is chosen to be $eaE_0/t_{\mr{hop}}=0.5$, while in panels (b1)-(b3), the laser frequency is set to $\hbar\omega /t_{\mr{hop}}=1.0$. Fitting dotted lines in panels (a3) and (b3) are proportional to $E_0^2/\omega^3$. 
    Other parameters are $t_{\rm hop}=1$, $\alpha_{\rm R}/t_{\mr{hop}}=0.1$, $B_x/t_{\mr{hop}}=0.1$, $\gamma/t_{\mr{hop}}=0.01$, and $\tau\simeq 14\hbar/t_{\rm hop}$. 
}
\label{fig:PulseomegaE0}
\end{figure}
Figure~\ref{Fig:Models_time}(a1)-\ref{Fig:Models_time}(a3) represents the slow precession motion of spins after the laser pulse passes. Each panel includes the numerically exact result of $s_{\rm slow}^\alpha(t)$ and the spin moments $\langle s^\alpha\rangle_t$ computed by the effective Hamiltonians $H_{\mr{FE}}^{\mr{Pulse}}(t)$ and $H_{\mr{FE}2}^{\mr{Pulse}}(t)$. 
As we discussed in Sec.~\ref{sec:Pulse-Prece}, $s_{\rm slow}^{x,y}(t)$ exhibits the laser-pulse driven precession mode in the $S^x$-$S^y$ plane, while $s_{\rm slow}^{z}(t)$ represents the slow oscillation after a small magnetization growth driven by the pulse.
The three Figures~\ref{Fig:Models_time}(a1)-\ref{Fig:Models_time}(a3) show that the Hamiltonian $H_{\mr{FE}}^{\mr{Pulse}}(t)$ well describes all the components of slow spin motion although the amplitudes of $\langle s^{y,z}\rangle_t$ somewhat deviates from the exact results of $s_{\rm slow}^\alpha(t)$. They also tell us that 
even a simpler Hamiltonian $H_{\mr{FE}2}^{\mr{Pulse}}(t)$ can reproduce the precession of $\langle s^{y,z}\rangle_t$ with the accurate frequency (although it cannot describe the $z$ component of spin). 

Figure~\ref{Fig:Models_time}(b1)-\ref{Fig:Models_time}(b3) show the Fourier transforms $s_{\mr{FT}}^\alpha(\Omega)$ for a right circularly polarized pulse. 
We plot results estimated by three methods: Numerically exact calculation by the GKSL equation, the numerical result based on $H_{\mr{FE}}^{\mr{Pulse}}(t)$ and that based on $H_{\mr{FE}2}^{\mr{Pulse}}(t)$. 
The curve of $s_{\mr{FT}}^\alpha(\Omega)$ estimated by $H_{\mr{FE}}^{\mr{Pulse}}(t)$ is shown to well agree with the exact result in the low-frequency regime of $\Omega<\omega$, while $H_{\mr{FE}2}^{\mr{Pulse}}(t)$ can describe the low-frequency dynamics for the $y$ and $z$ components of spin and cannot do the $x$ component. 
These results are consistent with the upper panels of Fig.~\ref{Fig:Models_time}(a). 
In particular, one finds from panels (b2) and (b3) that both models of Eqs.~(\ref{hhefPulse}) and (\ref{hhefPulse2}) extract the lowest-frequency peak structures of $s_{\mr{FT}}^{y,z}(\Omega)$. 

The broad peaks at $\Omega=\omega$, $2\omega$, and $3\omega$ in the numerically exact curves of Fig.~\ref{Fig:Models_time}(b) are a sort of the high harmonic generation induced by the laser pulse, and these high-frequency structures cannot be captured by the Floquet Hamiltonian.

We here define the characteristic frequency $\Omega_{\mr{slow}}^\alpha$ as the highest peak position of $s_{\mr{FT}}^\alpha(\Omega)$ that is smaller than the laser frequency $\omega$. 
The positions of $\Omega_{\mr{slow}}^\alpha$ are depicted in Fig.~\ref{Fig:Models_time}(b). 
The frequency $\Omega_{\mr{slow}}^y$ can be viewed as the frequency of the slow precession mode in Figs.~\ref{PulseMxLowhigh} and \ref{Fig:PulsePonchi}.
Figure~\ref{fig:PulseomegaE0} shows ${s_{\mr{FT}}^\alpha(\Omega_{\mr{slow}}^\alpha)}$ as a function of the AC electric field $E_0$ and the laser frequency $\omega$. It is found that 
${s_{\mr{FT}}^\alpha(\Omega_{\mr{slow}}^\alpha)}$ computed by combining the effective Hamiltonian (\ref{hhefPulse}) and the GKSL equation is in good agreement with the numerically exact result. 
This means that the Hamiltonian (\ref{hhefPulse}) works well to describe the dynamics that is slower than the laser frequency. Moreover, Figs.~\ref{fig:PulseomegaE0}(a3) and (b3) demonstrate that ${s_{\mr{FT}}^\alpha(\Omega_{\mr{slow}}^z)}$ obeys the line $\propto E_0^2/\omega^3$ like Eq.~(\ref{eq:mag_power}). 
It implies that the Floquet picture still survives even for the case of a few cycle laser pulse.

From the results of Figs.~\ref{Fig:Models_time} and \ref{fig:PulseomegaE0}, we conclude that the effective Hamiltonian for the laser-pulse driven systems, Eqs.~(\ref{hhefPulse}) and (\ref{hhefPulse2}), can capture the lower-frequency dynamics and the pulse-driven IFE can be viewed as a short-time FE.

\begin{figure}[t]
    \centering
    \includegraphics[width=0.45\textwidth]{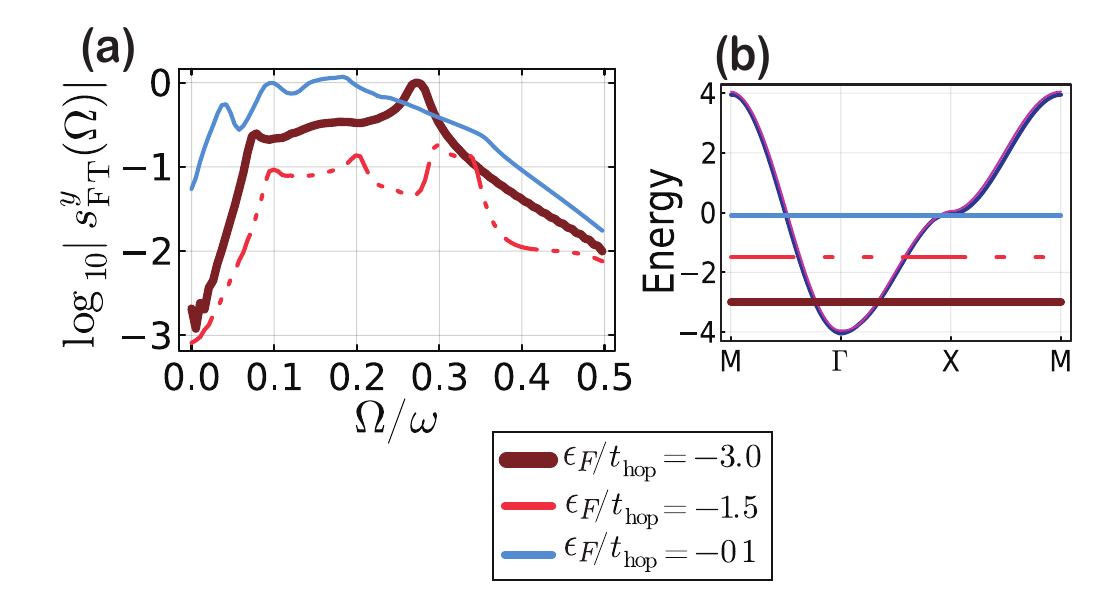}
\caption{(a) Fermi-energy ($\epsilon_F$) dependence of $\log_{10}|{s_{\mr{FT}}^y(\Omega)}|$ and the characteristic frequency $\Omega_{\mr{slow}}^y$ in the laser-pulse driven system of $H_{\rm FM}(t)$ with $t_{\rm hop}=1$, $\alpha_{\rm R}=0.1$ and $B_x=0.1$ at $T=0$. Three curves are the results of $\epsilon_F/t_{\rm hop}=-0.1$, $-1.5$ and $-3.0$. The highest peak position in $\log_{10}|{s_{\mr{FT}}^y(\Omega)}|$ corresponds to $\Omega_{\mr{slow}}^y$. 
(b) Lines of $\epsilon_F/t_{\rm hop}=-0.1$, $-1.5$ and $-3.0$ in the energy band of $H_{\rm FM}$ with $t_{\rm hop}=1$, $\alpha_{\rm R}=0.1$ and $B_x=0.1$. Other parameters are the same as those in Fig.~\ref{Fig:Models_time}. 
}
\label{fig:PrecessionFreq}
\end{figure}

Before ending this subsection, we shortly comment on the frequency $\Omega_{\mr{slow}}^y$ of the pulse-driven precession mode. As we mentioned, in real experiments of laser-pulse IFE, the precession frequency is given by the spin-wave frequency of the target magnetic material. However, our mean-field model does not include such a spin-wave mode, and the precession frequency $\Omega_{\mr{slow}}^y$ is expected to be associated with some parameters of $H_{\rm FM}$. Through the numerical computation, we verify that $\Omega_{\mr{slow}}^y$ strongly depends on 
the Fermi energy $\epsilon_F$ and the magnetic field $B_x$ (equivalently magnetization), whereas it is stable against a moderate change of laser-pulse width $\tau$, laser-field $E_0$, and laser frequency $\omega$. Figure~\ref{fig:PrecessionFreq} clearly shows that $\log_{10}|{s_{\mr{FT}}^y(\Omega)}|$ and $\Omega_{\mr{slow}}^y$ change depending on the value of $\epsilon_F$. 
This result is reasonable because the laser-driven effective Zeeman interaction $B_{\rm eff}({\bm k})$ depends on the wave vector 
$\bm k$ [see Eqs.~(\ref{SzCW}) and (\ref{Eq:hhEFCW})]. The $\epsilon_F$ dependence of $\Omega_{\mr{slow}}^y$ might be observed in a laser-pulse driven IFE for a nearly paramagnetic metal with a small magnetic moment.

\subsection{\label{sec:Pulse-Effect}Importance of Relaxation}
\begin{figure*}[t]
    \centering
    \includegraphics[width=\textwidth]{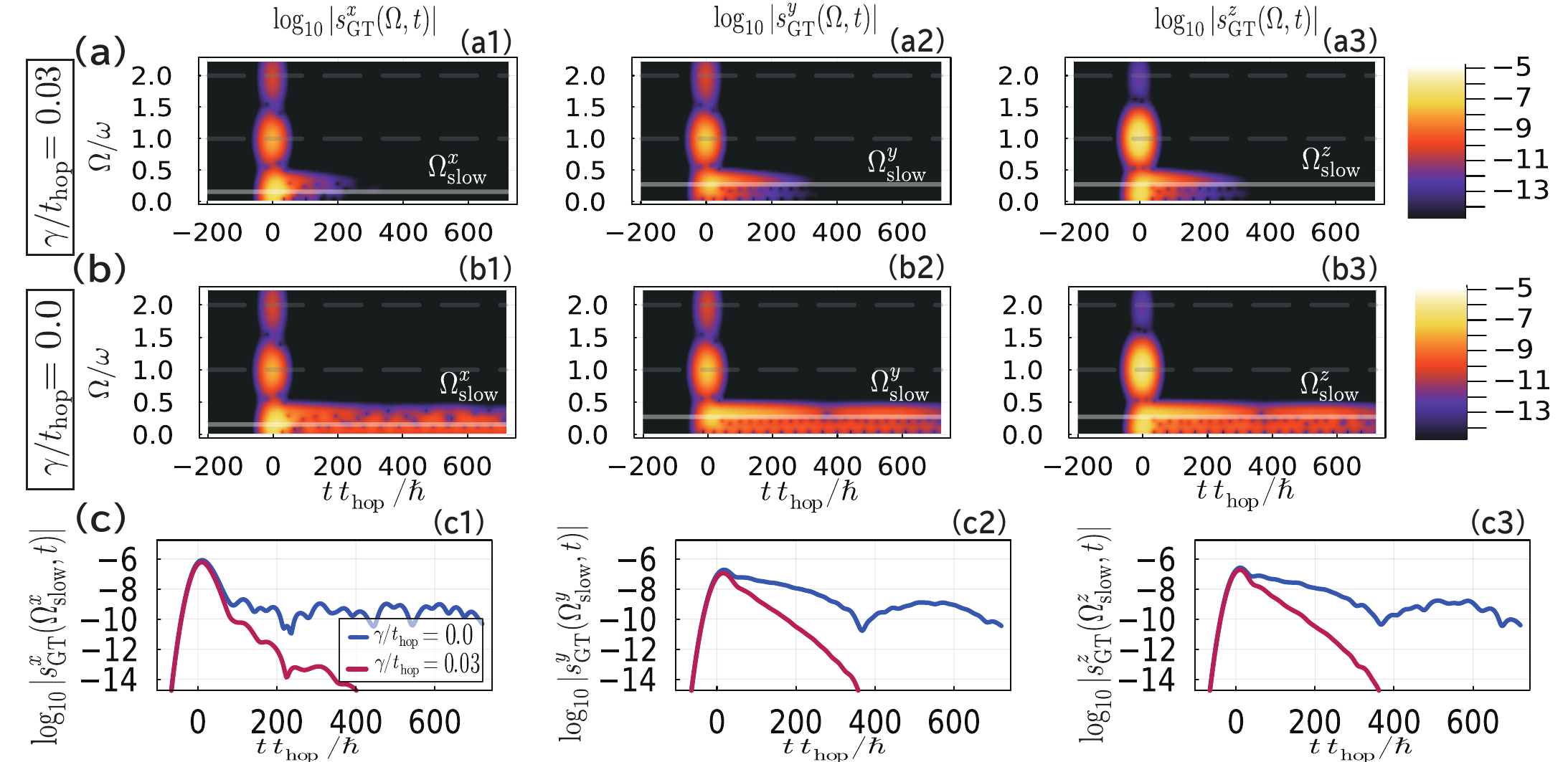}
    \caption{
[(a),(b)] Gabor transformation of $\ev {s^\alpha}_t$, $s_{\mr{GT}}^\alpha(\Omega, t)$, as a function of time $t$ and the observed frequency $\Omega$. The magnitude of $\log_{10}|s_{\mr{GT}}^\alpha(\Omega,t)|$ is color plotted. Panels (a) and (b) respectively correspond to the cases with a finite relaxation rate ($\gamma/t_{\mr{hop}}=0.03$) and without dissipation ($\gamma/t_{\mr{hop}}=0$). 
Other parameters are chosen to be $t_{\rm hop}=1$, $\alpha_{\rm R}/t_{\mr{hop}}=0.1$, $B_x/t_{\mr{hop}}=0.1$, $eaE_0/t_{\mr{hop}}=0.5$, $\hbar\omega /t_{\mr{hop}}=1$ and $T=0$. 
(c) $\log_{10}|s_{\mr{GT}}^\alpha(\Omega,t)|$ at $\Omega=\Omega_{\rm slow}^\alpha$ for $\gamma/t_{\mr{hop}}=0.03$ (red line) and $\gamma/t_{\mr{hop}}=0$ (blue line). 
    }
\label{fig:PreferGam}
\end{figure*}
Finally, we discuss how important the dissipation terms of the GKSL equation is when we consider the laser-pulse driven IFE. To this end, we introduce the Gabor transformation 
\begin{align}
{s_{\mr{GT}}^\alpha(\Omega,t)}=
\frac{t_{\rm hop}}{\hbar}\int^{\infty}_{-\infty}dt_s e^{-(t_s-t)^2/\xi^2}\Delta\ev{s^\alpha}_{t_s} e^{-i\Omega t_s},
\end{align}
where $\xi=0.1 \hbar/t_{\mr{hop}}$. This quantity tells us how the frequency components of $\ev{s^\alpha}_{t}$ are distributed at each time. 
This is also referred to as short-time Fourier transform. 

Figures~\ref{fig:PreferGam} (a1)-~\ref{fig:PreferGam}(a3) and ~\ref{fig:PreferGam}(b1)-~\ref{fig:PreferGam}(b3) show the above Gabor transforms for the case of laser pulse as a function of the frequency $\Omega$ and time $t$. Figures \ref{fig:PreferGam}(a1)-\ref{fig:PreferGam}(a3) are the results by the GKSL equation for the model~(\ref{Eq:timedepH_pulse}) with a finite dissipation strength $\gamma=0.03\:t_{\rm hop}$, while Figs.~\ref{fig:PreferGam}(b1)-~ref{fig:PreferGam}(b3) correspond to the results without dissipation term ($\gamma=0$). 
The results of $\gamma=0$ are shown to be almost the same as those of $\gamma=0.03t_{\rm hop}$ in the higher-frequency regime ($\Omega\gg\Omega_{\rm slow}^\alpha$). 
On the other hand, these figures also show that in the dissipationless case ($\gamma=0$), the low-frequency ($\Omega\sim \Omega_{\rm slow}^\alpha$) dynamics including the precession in Fig.~\ref{PulseMxLowhigh} survives for a long time, whereas the precession gradually relaxes in the case of $\gamma\neq 0$, i.e., we have a finite spin relaxation time. 
From Figs.~\ref{fig:PreferGam}(c1)-\ref{fig:PreferGam}(c3), we find that in the dissipationless system, the amplitude of the pulse-driven precession mode still remains at least in $t\sim 800\hbar/t_{\mr{hop}}$. 

The typical relaxation time of electron spins is in the range of \SI{1}{ps}--\SI{1}{ns}~\cite{kirilyuk2010,beaurepaire1996, koopmans2000, oshikawa2002, zutic2004, lenz2006, white2007,vittoria2010, furuya2015, mashkovich2019, tzschaschel2019} in magnetic materials.  
Therefore, the 
never-ending tails of the low-frequency regime in Fig.~\ref{fig:PreferGam}(b1)-(b3) are non realistic and it indicates the importance of taking the dissipation effect into account when we consider the laser-pulse driven dynamics. 

\begin{figure*}[t]
    \centering
    \includegraphics[width=\textwidth]{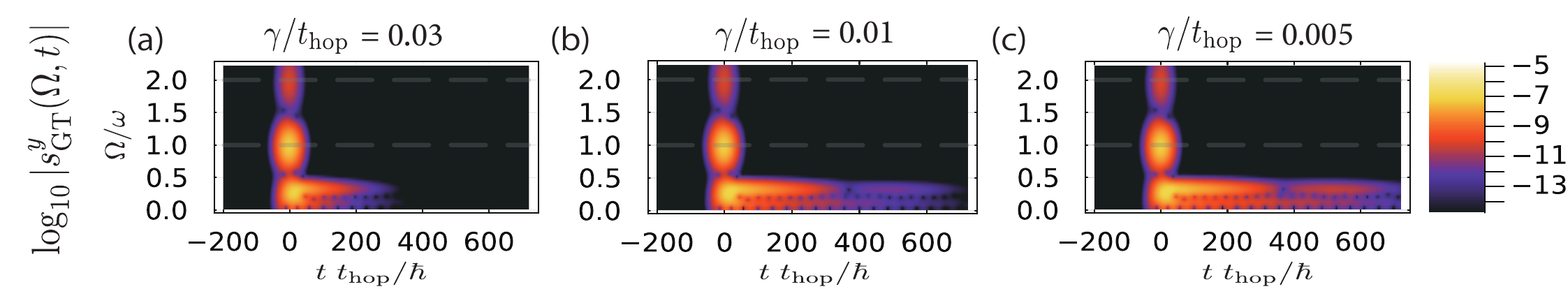}
    \caption{Dissipation-strength ($\gamma$) dependence of the Gabor transform of spin, ${s_{\mr{GT}}^y(\Omega,t)}$, in the laser-pulse driven ferromagnetic state in $H_{\rm FM}(t)$. Panels (a), (b) and (c) respectively correspond to the results of $\gamma/t_{\rm hop}=0.03$, $0.01$, and $0.005$. The other parameters are all the same as those of Fig.~\ref{fig:PreferGam}. As expected, the low-frequency mode ($\Omega<\omega$) survives longer with $\gamma$ decreasing.
    }
    \label{fig:GammaDep}
\end{figure*}

We note that the dissipation parameter $\gamma=0.03 t_{\rm hop}$ used in Fig.~\ref{fig:PreferGam} corresponds to the time scale $\hbar/\gamma\sim {\cal O}(10)$fs for $t_{\rm hop}\sim $\SI{1}{eV} and is too larger to obtain a typical relaxation time of the spin precession motion. However, the key point is that (as we mentioned) the spin relaxation time can be ``finite'' by taking account for the dissipation effect with the help of the GKSL equation, whereas a never-ending precession survives within the dissipationless Schr\"{o}dinger equation. One can easily expect that if we make the value of $\gamma$ sufficiently small, the spin relaxation time becomes long and approaches to a typical value. 
To clearly verify this expectation, we depict the $\gamma$ dependence of $s_{\mr{GT}}^y(\Omega,t)$ in Fig.~\ref{fig:GammaDep}. As expected, the life time of the low-frequency precession is shown to grow with $\hbar/\gamma$ increasing. Therefore, the GKSL equation approach can control the relaxation time of the pulse-driven precession with tuning the phenomenological parameter $\gamma$. If we further introduce a $\bm k$ or $\Omega$ dependence of $\Gamma_{ij}^{\bm k}$ in Eq.~(\ref{eq:GKSL2}), it would be possible to control the relaxation times of different observables.

\section{\label{sec:Comp} Comparisons}
As we mentioned in the Introduction, the IFE has long been explored and hence there are a lot of papers associated with IFEs. Here, we shortly compare our present theory and some of the previous analyses.

A standard method of accurately describing laser-driven dynamics in many-electron systems is to (numerically) solve the Schr\"odinger equation for isolated systems~\cite{tanaka2020, amano2022, takayoshi2014, takayoshi2014a} (although it can be applied to only small-size systems).
This method has been often applied in theoretical works of photo-induced phenomena in solids, especially, in correlated systems, and has captured their short-time evolution. 
On the other hand, the GKSL equation approach used in this study can capture the dissipation effect, which is inevitable in experiments, and hence it can describe both short- and long-time evolution of the laser-driven phenomena. This study focuses on IFEs and their dissipation effects in metallic systems, while Refs.~\cite{ikeda2020, ikeda2021} have analyzed dissipation effects of THz-laser driven IFEs in magnetic insulators.

In addition to the numerical methods, recently, Floquet theory techniques~~\cite{eckardt2015, eckardt2017,oka2019, Sato2021} have begun to be applied to IFEs~\cite{Sato2021,tanaka2020, banerjee2022}. In particular, the Floquet Hamiltonian is useful to understand the essential picture of IFE~\cite{tanaka2020, banerjee2022}. We have indeed used the effective Hamiltonian in this work. However, we note that (as discussed in Sec.~\ref{sec:Pulse}) the Floquet Hamiltonian is not enough to quantitatively understand the time evolution of physical quantities. Estimating the time evolution during laser application is necessary for an accurate prediction and an explanation for experimental results of IFEs.

Perturbation approaches with respect to laser fields~\cite{pershan1966, battiato2014, berritta2016} are powerful to obtain the lower-order effects of laser. In particular, they enable one to obtain analytical expressions for laser-driven quantities. 
It is difficult to obtain such analytical results from the numerical computation based on Schr\"odinger or GKSL equations, whereas the numerical approaches can capture non-perturbative effects of laser (see, e.g., Fig.~\ref{Fig:CW_omegaE0}).

Most of the above techniques basically start from a microscopic Hamiltonian. On the other hand, phenomenological theories have also been developed for IFEs. 
For instance, Ref.~\cite{dannegger2021} has developed a combination method consisting of the perturbation theory, the stochastic Landau-Lifshitz-Gilbert (LLG) equation~\cite{maekawa2017,kirilyuk2010} and the phenomenological multiple-temperature model~\cite{kirilyuk2010}. The perturbation theory (with the electron-band structure) is used to compute the instantaneous magnetic field driven by a circularly-polarized laser pulse, and then the LLG and multiple-temperature models compute the ultrafast magnetization dynamics generated by the instantaneous field, including laser-heating effects. This method effectively describes two different time-scale dynamics of photo-induced electron transitions and the correlated spin dynamics after laser pulse application. In particular, it is powerful to discuss laser-heating effects. However, we should note that such a phenomenology requires several fitting parameters. On the other hand, in this work, we have tried to develop a microscopic theory for IFE that relies as little on phenomenological parameters as possible. As a result, for instance, we have succeeded in describing the laser-pulse driven spin precession from the microscopic Hamiltonian (and a simple jump operator). 
It is generally important to develop the microscopic theory with reference to the phenomenology, to deeply understand IFEs.

\section{\label{sec:Con} Conclusions and discussions}
In this study, we have theoretically investigated the IFE driven by continuous (pulse) waves for paramagnetic (ferromagnetic) states in 2D Rashba electron models. 
The quantum master (GKSL) equation~\cite{lindblad1976,gorini1976,breuer2007,alicki1987} makes it possible to 
handle the spin dynamics driven by both pulse and continuous waves. 
Moreover, it also enables us to take the dissipation effects into account 
unlike the standard approach of the Schr\"{o}dinger equation. 
We emphasize that the dissipation plays the significant roles to realize the NESS in the case of CW (see Sec.~\ref{sec:Ness}) and to describe the spin dynamics in the case of laser pulse (see Sec.~\ref{sec:Pulse}). Through the comprehensive analyses in Secs.~\ref{sec:Ness} and \ref{sec:Pulse}, 
we have succeeded in revealing some fundamental properties of the laser-driven NESS and laser-pulse driven precession, starting from the microscopic Hamiltonians.

In Sec.~\ref{sec:Ness}, we have investigated the NESS that arises due to the balance between the energy injection by CW laser and the energy dissipation. 
We demonstrate that the laser-driven magnetization and its nature can be captured by the Floquet theory for dissipative systems~\cite{ikeda2020, ikeda2021} in the high-frequency regime. 
We analytically and numerically prove that the power law of magnetization, $\langle\langle s^z\rangle\rangle\propto E_0^2/\omega^3$, holds in the NESS. 
With the GKSL equation, we predict the $T$ and $\gamma$ dependence of IFE in a quantitative level. For example, for a typical value of $t_{\rm hop}$, the laser-driven magnetization is shown to remain large enough even when $T$ is as high as room temperature. 

Section~\ref{sec:Pulse} is devoted to the analysis of the short-laser-pulse-driven IFE. 
We have focused on ferromagnetic metal states by introducing a mean-field type Zeeman interaction and computed the pulse-driven ultrafast spin dynamics in the ferromagnetic state by using the GKSL equation. 
We find that a pulse-induced instantaneous magnetic field leads to a precession of the spin moment, which has been often observed in experiments of IFE. 
Furthermore, by introducing time-dependent effective Hamiltonians, we show that the precession can be understood from the Floquet theory perspective. 
We note that when even ``a few'' cycle pulse is applied, 
the Floquet picture is still useful to understand some essential properties of the laser-pulse-driven systems.  

Our results indicate that the GKSL equation~\cite{lindblad1976,gorini1976,breuer2007,alicki1987} and the Floquet theory for dissipative systems~\cite{ikeda2020, ikeda2021} 
are useful to deeply understand Floquet-engineering phenomena in solids including IFE, in which the dissipation effect is usually inevitable. 
On the other hand, (as we discussed in Sec.~\ref{sec:Pulse-Prece}), the pulse-driven precession with the spin-wave frequency cannot be reproduced within our free electron model. The development of a microscopic theory for such a spin-wave precession mode is an interesting future issue. 
The analysis of dissipation effects beyond the GKSL equation~\cite{passos2018, michishita2021, terada2024a} is also important in broad fields of non-equilibrium physics.

\begin{acknowledgments}
This work is supported by JSPS KAKENHI (Grant No. 20H01830 and No. 20H01849) and a Grant-in-Aid for Scientific Research 
on Innovative Areas ”Quantum Liquid Crystals” (Grant No. 19H05825) and 
“Evolution of Chiral Materials Science using
Helical Light Fields” 
(Grants No. JP22H05131 and No. JP23H04576) 
from JSPS of Japan.
\end{acknowledgments}

\appendix

\section{\label{sec:A-Table} Typical values of parameters}
When we theoretically study laser-driven systems, we should consider realistic values of many parameters. The number of them is generally much larger than that in equilibrium systems. 
Here, we prepare two tables, Tables
 ~\ref{Table1} and \ref{Table2}, in which typical values of important parameters are listed. 

\begin{table*}[t]
\caption{Several quantities that are used when we consider laser-driven electron systems. In the second column, we define typical quantities in the dimensionless fashion. The symbol ``dl'' means ``dimensionless''. In the third column, we show typical values of these quantities under the condition that the dimensionless parameters are fixed to $E_0^{(\mr{dl})}=1$, 
$\hbar\omega^{(\mr{dl})}=1$, $T^{(\mr{dl})}=1$, $t^{\mr(dl)}=1$ and $\gamma^{(\mr{dl})}=1$ and we set $t_{\rm{hop}}=$\SI{1}{eV} and $a=5$~\AA. See Ref.~\cite{tanaka2020}.}
\label{table:Params}
 \centering
  \begin{tabular}{lllllll}
   \hline
     parameter &  dimensionless parameter &  $E_0^{(\mr{dl})}=1,\;\hbar\omega^{(\mr{dl})}=1,\;T^{(\mr{dl})}=1,\;\gamma^{(\mr{dl})}=1,\;t^{\mr{(dl)}}=1$ \\
   \hline \hline
    Strength of laser electric field &$E_0^{(\mr{dl})}= ea E_0/t_{\mr{hop}}$ & $E_0\simeq$ \SI{20}{MVcm^{-1}} \\
    Strength of laser magnetic field &$B_0^{(\mr{dl})}
    =g\mu_B B_0/t_{\mr{hop}}=g\mu_B(E_0/c)/t_{\mr{hop}}$\;\;\;\;&  $ B_0\simeq$\SI{6.67}{T} \\
    laser angular frequency &$\hbar\omega^{(\mr{dl})}= \hbar \omega /t_{\mr{hop}}$&  $\frac{ \omega}{2\pi}\simeq 2.42\times 10^2${THz}\\
    Temperature \;\;\;\;&$k_BT^{(\mr{dl})}= \;k_B T/t_{\mr{hop}}$&  $ T\simeq$ $1.16\times 10^4$ {K}\\
   strength of dissipation &$\gamma^{(\mr{dl})}= \gamma/t_{\mr{hop}}$ & $\gamma=$\SI{1}{eV} \\
    time &$t^{(\mr{dl})}= {t}\;{t_{\mr{hop}}}/{\hbar}$ & ${t}\simeq$\SI{0.66}{fs}  \\
  \hline
  \end{tabular}
  \label{Table1}
\end{table*}
\begin{table*}[t]
\caption{Laser energy flux for reference field strengths. See Ref.~\cite{ikeda2019}.
}
\label{table:Param}
 \centering
  \begin{tabular}{lll}
   \hline
     \:  &  $E_0=$\SI{1}{MVcm^{-1}} \\
   \hline \hline
   Strength of laser magnetic field ($B_0$) & \SI{0.33}{T}\\
    Laser energy flux ($I$) & \SI{1.3}{GWcm^{-2}}\\
  \hline
  \end{tabular}
  \label{Table2}
\end{table*}

\section{\label{sec:A-GKSLBloch} Relation between GKSL and Bloch Equations}
In this study, we have treated dissipation effects by using the GKSL equation. Below, we explain that the GKSL equation encompasses the (optical) Bloch equation \cite{haug2008}, 
which has been often used to describe photo-induced dynamics in semiconductors.
We assume that the system is given by a two-level quantum model, whose $2\times 2$ Hamiltonian denotes $H$. The GKSL equation for the density matrix $\rho$ is written as follows:
\begin{equation}
\label{GKSL}
\dv{{\rho}}{t}=-i\qty[{H},{\rho}]+\sum_{m}\qty({L}_m{\rho} {L}_m^\dagger-\frac{1}{2}\qty{{L}_m^\dagger {L}_m,\rho}),
\end{equation}
where $L_m$ is a jump operator describing a dissipation process. We note that the dissipation term in Eq.~(\ref{GKSL}) can be re-expressed as Eq.~(\ref{eq:GKSL2}) by changing the definition and normalization of $L_m$. 
In the two-level system, arbitrary operator $A$ is given by $A=\frac{1}{2}\sum_{\alpha=0,x,y,z}A_{\alpha}\sigma_\alpha$, where $\sigma_{x,y,z}$ are Pauli matrices and $\sigma_0$ is the $2\times 2$ unit matrix. 
The density matrix $\rho$ is hence expressed as 
$\rho=\frac{1}{2}\sum_{\alpha=0,x,y,z}\rho_{\alpha}\sigma_\alpha$ and the determination of the density matrix is equivalent to giving the vector of three coefficients ${\bm \rho}=(\rho_x,\rho_y,\rho_z)$. 
Similarly, the Hamiltonian $H$ and each jump operator $L_m$ may be respectively defined by three-component vectors ${\bm H}=(H_x,H_y,H_z)$ and  ${\bm L}_m=(L_{m,x},L_{m,y},L_{m,z})$. Using these instruments, one can exactly map the GKSL equation to the following differential equation: 
\begin{align}
    \label{bloEq}
   \dv{t}\mqty( \rho_x\\ \rho_y\\ \rho_z)
        =&\bm H\times\bm \rho +
       \sum_m \mqty(
                -\tau^{x}_m & \tau^{xy}_m &  \tau^{xz}_m \\
                 \tau^{yx}_m & -\tau^{y}_m &  \tau^{yz}_m \\
                \tau^{zx}_m &  \tau^{zy}_m & -\tau^{z}_m 
                )\mqty( \rho_x\\ \rho_y\\ \rho_z)\no\\
        &+i
       \sum_m \mqty(\bm L_m \times\bm L^\dagger_m)/2,
\end{align}
where $\tau^{i}_m=(|L_{m,i+1}|^2 +|L_{m,i+2}|^2)/2$ 
(here, $x$, $y$, and $z$, respectively, read as 1, 2, and 3 mod 3) and $\tau^{ij}_m=(L_{m,j}{L_{m,i}}^*+{L_{m,j}}^*L_{m,i})/4$ for $i\neq j$. The trace conservation of $\rho$ leads to the traceless nature of ${\rm Tr}[L_m]=0$.

The jump operators are here determined such that $\rho(t)$ approaches to an equilibrium state of the time-independent Hamiltonian $H_{\rm stat}$: The full  Hamiltonian is given by $H(t)=H_{\rm stat}+\delta H(t)$. 
To quantitatively discuss the jump operators, we prepare the eigenstates of $H_{\rm stat}$.  
Two energy eigenstates of $H_{\rm stat}$ are defined by $|E_1\rangle$ and $|E_2\rangle$, whose eigenenergies $E_{1,2}$ satisfy $E_1<E_2$. In this setup, the density matrix
\begin{align}
\rho_{\rm gs}=|E_1\rangle\langle E_1|=
\mqty(
1& 0\\
0& 0 )
\end{align}
corresponds to the ground state. 

To consider the relation between the GKSL and Bloch equations, we somewhat restrict the form of the jump operators: Each $L_m$ is assumed to be proportional to $\sigma_z$, $\sigma_+=(\sigma_x+i\sigma_y)/2$ or $\sigma_-=(\sigma_x-i\sigma_y)/2$. For example, we do not consider the case where $L_m$ is given by a linear combination of $\sigma_z$ and $\sigma_{\pm}$. 
Under this constraint, the matrices in the second term of Eq.~(\ref{bloEq}) possesses only diagonal components $\tau_m^i$. Similarly, the vector $\bm L \times\bm L^\dagger$ can have only the $z$ component. 

\subsection{\label{sec:A-GKSLBloch-NonLz} 
Absence of \texorpdfstring{$\sigma_z$}{sigmaz}}
Here, we consider the case where jump operators do not include any diagonal component. Two jump operators are defined by 
\begin{align}
L_1=&\sqrt{\Gamma_{12}}\sigma_+=\sqrt{\Gamma_{12}}\ket{E_1}\bra{E_2},\nonumber\\
L_2=&\sqrt{\Gamma_{21}}\sigma_-=\sqrt{\Gamma_{21}}\ket{E_2}\bra{E_1}.
\label{eq:trans}
\end{align}
This setup corresponds to the GKSL equation we have defined in Sec.~\ref{sec:Mode-GKSL}: 
The relation between the jump operator in Sec.~\ref{sec:Mode-GKSL} and that in Eq.~(\ref{eq:trans}) is given by $L_{12}=L_1/\sqrt{\Gamma_{12}}$ and $L_{21}=L_2/\sqrt{\Gamma_{21}}$. 
The equation of motion for the density matrix is 
\begin{align}
   \dv{\rho_{x}}{t}
      &=\qty[\bm H\times \bm \rho]_x
      -\frac{\rho_x}{T_{\perp}}\\
    \dv{\rho_y}{t}
      &=\qty[\bm H\times \bm \rho]_y
      -\frac{
       \rho_{y}}{T_{\perp} }\\
    \dv{\rho_{z}}{t}
      &=\qty[\bm H\times \bm \rho]_z
      -\frac{\rho_{z}-\frac{\Gamma_{12}-\Gamma_{21}}{\Gamma_{12}+\Gamma_{21}}}{T_{\parallel}},
      \label{eq:Blochz}
\end{align}
where $1/T_\perp=(\Gamma_{12}+\Gamma_{21})/2$ and $1/T_\parallel=\Gamma_{12}+\Gamma_{21}$. 
This equation is nothing but the same form as the Bloch equation. 
If we regard the vector $\bm \rho$ as a classical spin, $T_\parallel$ and $T_\perp$ may be respectively interpreted as the longitudinal and transverse relaxation times. 
Like Sec.~\ref{sec:Mode-GKSL}, if $\Gamma_{ij}$ satisfy the detailed balance condition
\begin{align}
    \Gamma_{ij}=\frac{\gamma\exp({-\beta E_i})}{\exp({-\beta E_i})+\exp({-\beta E_j})}\;\;(i\neq j),
\end{align}
the system relaxes to the equilibrium state of the time-independent Hamiltonian. In fact, 
we find that the factor in the final term of Eq.~(\ref{eq:Blochz}) satisfies 
\begin{align}
\frac{\Gamma_{12}-\Gamma_{21}}{\Gamma_{12}+\Gamma_{21}}=\frac{\exp({-\beta E_1})-\exp({-\beta E_2})}{\sum_{i=1,2}\exp({-\beta E_i})}=\ev {\rho_z}_{\rm eq},
\end{align}
where $\langle\cdots\rangle_{\rm eq}$ denotes the expectation value of an equilibrium state. 
Moreover, one finds $T_{\perp}=2T_{\parallel}$. Namely, the use of Eq.~(\ref{eq:trans}) corresponds to the Bloch equation under a special condition that the relaxation times satisfy $T_{\perp}=2T_{\parallel}$.  
At $T=0$ ($\beta \rightarrow \infty$), $\Gamma_{21}\rightarrow 0$ and $\ev{\rho_z}_{\rm eq}=1$.

\subsection{\label{sec:A-GKSLBloch-Lz} 
Existence of \texorpdfstring{$\sigma_z$}{sigmaz}}
In addition to $L_{1,2}$, we introduce another jump operator with a diagonal component:
\begin{align}
    L_{3}=\sqrt{\frac{\gamma_\perp}{2}}\,\,\sigma_{z}.
\end{align}
From Eq.~(\ref{bloEq}), the GKSL equation with three jump operators $L_{1,2,3}$ are expressed as  
\begin{align}
   \dv{\rho_x}{t}
      &=\qty[\bm H\times \bm \rho]_x
      -\left(\frac{1}{T_{\perp1}}
      +\frac{1}{T_{\perp2}}\right)\rho_x\\
    \dv{\rho_y}{t}
      &=\qty[\bm H\times \bm \rho]_y
      -\left(\frac{1}{T_{\perp1}}
      +\frac{1}{T_{\perp2}}\right)\rho_y\\
    \dv{\rho_z}{t}
      &=\qty[\bm H\times \bm \rho]_z
      -\frac{\rho_z-\frac{\Gamma_{12}-\Gamma_{21}}{\Gamma_{12}+\Gamma_{21}}}{T_{\parallel}},
\end{align}
where $1/T_{\perp1}=(\Gamma_{12}+\Gamma_{21})/2=\gamma/2$ and $1/T_{\perp2}=\gamma_\perp$. This is also equivalent to a Bloch equation. 
The longitudinal relaxation time is $T_\parallel=1/\gamma$, while the transverse relaxation time is given by $T_\perp=(1/T_{\perp1}+1/T_{\perp2})^{-1}=2/(\gamma+2\gamma_\perp)$. 
That is, one sees that the additional jump operator $L_3$ contributes to only transverse relaxation process and it is not sufficient to make the system relax to an equilibrium state. This is because $L_3$ does not include any transition between the ground state $|E_1\rangle$ and the excited one $|E_2\rangle$. We can control the magnitude and ratio of $T_\perp$ and $T_\parallel$ by tuning the dissipation strength of jump operators, $\gamma$ and $\gamma_\perp$. This control is impossible when we have only $L_{1,2}$. We note that $2T_\parallel\geq T_\perp$ holds in this Bloch (or GKSL) equation \cite{kimura2002,kimura2017}.  


It is noteworthy that recently there have been a few advancements in theoretical studies that argue some issues about the relaxation-time approximation in the GKSL equation~\cite{passos2018, michishita2021, terada2024a}.

\section{\label{sec:A-SolNESS} Density matrices in NESS}
In this appendix, we consider the density matrices of the laser-driven NESS in GKSL equations. 
We focus on the GKSL equation for laser-driven time-periodic systems with a static jump operator like Eqs.~(\ref{eq:GKSL1}) and (\ref{eq:GKSL2}):
\begin{align}
\label{GKSLstatic}
\dv{\rho(t)}{t}=\LL \rho(t) = -i[H(t),\rho(t)] + \mathcal{D}(\rho(t)),
\end{align}
where $\rho(t)$ is the density matrix, $H(t)=H(t+T_\omega)$ is the time-periodic Hamiltonian, and $\mathcal{D}(\rho(t))$ is the dissipation part with a static jump operator $L_m$. For the dissipative quantum system, we can apply the Floquet high-frequency expansion~\cite{ikeda2020,ikeda2021} like the case of Schr\"{o}dinger equations for isolated systems. To this end, we divide the time evolution operator into three parts as follows:
\begin{align}
    U(t,0)=e^{\mathcal{G}(t)}\rho^{t\LL_{\rm{eff}}}e^{-\mathcal{G}(0)},
\end{align}
where $U(t,0)$ is the time evolution operator, $\mathcal{G}(t)=\mathcal{G}(t+T_\omega)$ is the micromotion operator describing the faster dynamics than the laser period $T_\omega=2\pi/\omega$, and $\LL_{\rm{eff}}$ is the time-independent Lindbladean describing the slow dynamics. Via the high-frequency expansion for $\LL_{\rm{eff}}$ and $\mathcal{G}(t)$, we obtain the following effective equation of motion for the slow dynamics~\cite{ikeda2020}: 
\begin{align}
\label{GKSLeff}
\dv{\rho(t)}{t}=\LL_{\rm{eff}} \rho(t) = -i[H_{\mr{FE}},\rho(t)] + \mathcal{D}(\rho(t)),
\end{align}
where time-independent Flouqet Hamiltonian $H_{\mr{FE}}$ is given by 
$H_{\mr{FE}}=H_0 +\sum_n\frac{[H_{-n}.H_n]}{n\omega}+\order{\omega^{-2}}$ and 
$H_n$ is defined from the Fourier transform of $H(t)$: $H(t)=\sum_{n}H_ne^{-in\omega t}$ ($n\in \mathbb{Z}$). 
Here, we define 
\begin{align}
    \tilde{\rho}(t)=e^{t\LL_{\rm{eff}}}e^{-\mathcal{G}(0)}\rho(0)
\end{align}
and $\tilde{\rho}_{\infty}={\rm lim}_{t\to\infty}\tilde{\rho}(t)$. 
Using them, the density matrix for the NESS is given by 
\begin{align}
\rho_{\mr{NESS}}(t)=e^{\mathcal{G}(t)}\tilde{\rho}_{\infty}, 
\end{align}
and we find that $\tilde{\rho}(t)$ satisfies 
\begin{align}
\dv{\tilde{\rho}(t)}{t}
=\LL_{\rm{eff}}\tilde{\rho}(t).
\label{eq:NESS_Static}
\end{align}
Because $\mathcal{G}(t)$ gives an oscillating factor to the density matrix, the main time-independent nature of the NESS is written in $\tilde{\rho}_{\infty}$. From Eq.~(\ref{eq:NESS_Static}), we see that $\tilde{\rho}_{\infty}$ is the solution of 
\begin{align}
\LL_{\rm{eff}}\tilde{\rho}(t)=0.
\label{eq:NESS_Sol}
\end{align}
Below we will explain the explicit form of $\tilde{\rho}_{\infty}$ in some representative setups.

\subsection{\label{sec:A-SolNESS-eq}\texorpdfstring{$H_0=H_{\rm stat}$}{H0=H}}
In laser-driven systems, the Hamiltonian is generally given by
\begin{align}
\label{eq:GenericH}
    H(t)=H_{\rm stat}+\delta H(t),
\end{align}
where $H_{\rm stat}$ is the time-independent part and $\delta H(t)=\delta H(t+T_{\omega})$ is the time-dependent periodic part. 
First, we consider the case where 
\begin{align}
    H_0=H_{\rm stat}.
    \label{eq:H0=H}
\end{align}
This condition often holds in periodically driven systems. In addition, we assume that jump operators 
are given by $L_{ij}=\ket{E_i}\langle E_j|$ and the corresponding coupling constants $\Gamma_{ij}$ satisfy the detailed balance condition,
\begin{align}
    &\Gamma_{ji}e^{-\beta E_i}=\Gamma_{ij}e^{-\beta E_j}\;\;\;\;\;\;(i\neq j),\nonumber\\
    &\Gamma_{ii}=0  
\label{eq:Gamma}
\end{align}
such that for $\delta H(t)=0$, the system approaches to the equilibrium state of $H_{\rm stat}$. Here, $E_i$ is the $i$-th eigenenergy of $H_{\rm stat}$ and $|E_i\rangle$ is the corresponding eigenstate. The solution of the NESS under the condition of Eqs.~(\ref{eq:H0=H}) and (\ref{eq:Gamma}) is given in Ref.~\cite{ikeda2020}. For simplicity, we restrict ourselves to the non-degenerate case: $E_i\neq E_j$ for $i\neq j$. 
Here, we shortly review the result of Ref.~\cite{ikeda2020}. 

To obtain the density matrix of the NESS, it is convenient to divide $\tilde{\rho}_{\infty}$ into the diagonal part 
$\tilde{\rho}_{\infty}^{(d)}$ and the off-diagonal one $\tilde{\rho}_{\infty}^{(od)}$ as follows:
\begin{align}
    \tilde{\rho}_{\infty}^{(d)}  := \sum_{k} \rho^{kk}\ket{E_k}\langle E_k|, &&
    \tilde{\rho}_{\infty}^{(od)} := \sum_{k,l(l\neq k)} \rho^{kl}\ket{E_k}\langle E_l|,
\end{align}
where $\tilde{\rho}_{\infty}=\tilde{\rho}_{\infty}^{(d)}+\tilde{\rho}_{\infty}^{(od)}$. 
Equation~(\ref{eq:NESS_Sol}) means $\bra{E_n}\LL_{\rm{eff}}{\tilde \rho_\infty}\ket{E_m}=0$. Focusing on the dissipation part, we have
\begin{align}
\label{Drho}
&\bra{E_n}\mathcal{D}(\tilde\rho_\infty)\ket{E_m}
\\
&=\sum_{i}\qty(\Gamma_{ni}\rho^{ii}-\Gamma_{in}\rho^{nn})\delta_{nm}
-\gamma_{nm}\rho^{nm}(1-\delta_{nm}),\nonumber
\end{align}
where $\gamma_{nm}=\frac{1}{2}\sum_{i} (\Gamma_{in}+\Gamma_{im})$. Secondly, considering the commutator part $-i\bra{E_n}[H_{\mr{FE}},{\tilde{\rho}_{\infty}}]\ket{E_m}$, we obtain 
\begin{align}
\label{eq:rhoOD}
&-i(E_n-E_m)\rho^{nm}
-i(\rho^{mm}-\rho^{nn})\bra{E_n}\Delta H_{\mr{FE}}\ket{E_m}\nonumber\\
&-i\bra{E_n}[\Delta H_{\mr{FE}},{\tilde{\rho}_{\infty}}^{(od)}]\ket{E_m}
-\gamma_{nm}\rho^{nm}=0,\,\,\,\,\,\,\,\,(m\neq n)
\\
&-i\bra{E_n}[\Delta H_{\mr{FE}},{\tilde{\rho}_{\infty}}^{(od)}]\ket{E_n}
+\sum_i(\Gamma_{ni}\rho^{ii}-\Gamma_{in}\rho^{nn})
=0,\nonumber\\
&\,\,\,\,\,\,\,\,(m=n)
\label{eq:rhoD1}
\end{align}
where $\Delta H_{\rm FE}=H_{\rm FE}-H_0$. 
By using these equalities, we can obtain $\rho^{kl}$. From Eq.~(\ref{eq:rhoD1}), we find that the diagonal elements $\rho^{ii}$ satisfies
$\sum_i(\Gamma_{ni}\rho^{ii}-\Gamma_{in}\rho^{nn})=0$ in the $(1/\omega)^0$ order. Therefore, we have 
\begin{align}
\rho^{(d)}_\infty
=\rho_{\mr{can}}+\order{\omega^{-2}},
\end{align}
where $\rho_{\mr{can}}$ is the canonical distribution  
\begin{align}
{\rho}_{\mr{can}}=
\frac{\sum_k e^{-\beta E_k}\ket{E_k}\langle E_k|}
{\sum_{l}e^{-\beta E_l}}
\end{align}
Using this result, we also obtain the off-diagonal elements:
\begin{align}
\tilde{\rho}_{\infty}^{(od)}=\sigma_{\mr{FE}}
+\order{\omega^{-2}},
\end{align}
where $\sigma_{\mr{FE}}$ is defined as 
\begin{align}
\langle E_n|\sigma_{\mr{FE}}|E_m\rangle
=\frac{\langle E_n|\Delta H_{\mr{FE}}|E_m\rangle}{(E_n-E_m)-i\gamma_{nm}}
(\rho_{\mr{can}}^{n}-\rho_{\mr{can}}^{m}).
\end{align}
This off-diagonal part represents the Floquet engineering. 

\subsection{\label{sec:A-SolNESS-neq}\texorpdfstring{$H_0\neq H_{\rm stat}$}{H0 not equal}}
In the following, we consider the case where $H_0 \neq H_{\rm stat}$, while Eqs.~(\ref{eq:GenericH}) and (\ref{eq:Gamma}) hold. In this case, Eq.~(\ref{Drho}) still holds, whereas we have to slightly modify the calculation below Eq.~(\ref{Drho}). First, we divide $H_0$ into two parts as follows: 
\begin{align}
    H_0=H_0^{(0)}+H_0^{(1)},
    \label{eq:H0omega}
\end{align}
where $H_0^{(0)}$ and $H_0^{(1)}$ are respectively $\order{\omega^{0}}$ and $\order{\omega^{-1}}$. Moreover, we define 
\begin{align}
    \Delta H_{\rm FE}^{(1)}=H_{\rm FE}-H_0^{(0)}, 
    \label{eq:HFEomega}
\end{align}
as an extension of $\Delta H_{\rm FE}$. 
The Floquet Hamiltonian is given by 
$H_{\rm FE}=H_0^{(0)}+\Delta H_{\rm FE}^{(1)}$ and $\Delta H_{\rm FE}^{(1)}$ is $\order{\omega^{-1}}$. These new parameters are useful to estimate the density matrix in terms of the power $1/\omega$. 
The remaining task is to compute the matrix elements $\bra{E_n}\LL_{\rm{eff}}\tilde\rho_{\infty}\ket{E_m}$. 
The diagonal $(m=n)$ and off-diagonal $(m\neq n)$ elements are computed as
\begin{align}
&-i(\rho^{mm}-\rho^{nn})\bra{E_n}H_0^{(0)}+\Delta H_{\rm{FE}}^{(1)}\ket{E_m}\nonumber\\
&-i\bra{E_n}[H_0^{(0)}+\Delta H_{\mr{FE}}^{(1)},\tilde{\rho}_{\infty}^{(od)}]\ket{E_m}-\gamma_{nm}\rho^{nm}=0,\nonumber\\
&\,\,\,\,\,\,\,\,\,\,(m\;\neq \;n)\\
&-i\bra{E_n}[H_0^{(0)}+\Delta H_{\rm{FE}}^{(1)},\tilde{\rho}_{\infty}^{(od)}]\ket{E_n}
+\sum_i(\Gamma_{ni}\rho^{ii}-\Gamma_{in}\rho^{nn})\nonumber\\
&=0.\,\,\,\,\,\,\,\,\,\,(m=n)
\end{align}
Here we have used the Hermitian natures 
$({\rho^{mn}})^*=\rho^{nm}$ and $\bra{E_m}H_{0}\ket{E_n}^*=\bra{E_n}H_{0}\ket{E_m}$. 

For simplicity, below we restrict ourselves to two-level systems, in which indices $m$ and $n$ take only two values of 1 and 2. 
In such two-level systems, the above equations are reduced to 
\begin{align}
&\Big(i\bra{E_n}H_0^{(0)}
+\Delta H_{\rm{FE}}^{(1)}\ket{E_n}-i\bra{E_m}H_0^{(0)}+\Delta H_{\rm{FE}}^{(1)}\ket{E_m}\no\\
&+\gamma_{nm}\Big)\rho^{nm}
+i(\rho^{mm}-\rho^{nn})\Big[\bra{E_n}\Delta H_{\rm{FE}}^{(1)}\ket{E_m}\no\\
&+\bra{E_n} H_0^{(0)}\ket{E_m}\Big]
=0, \,\,\,\,\,\,\, (m\neq n)\\
&i\rho^{kn}[\bra{E_n}H_0^{(0)}
+\Delta H_{\rm{FE}}^{(1)}\ket{E_{k}}]\no\\
&-i\rho^{mn}[\bra{E_k}H_0^{(0)}+\Delta H_{\rm{FE}}^{(1)}\ket{E_n}]+(\Gamma_{nk}\rho^{kk}-\Gamma_{kn}\rho^{nn})\no\\
&=0. \,\,\,\,\,\,\, (m=n\neq k)
\end{align}
Solving these four equations, we can obtain all the matrix elements of $\rho^{11}$, $\rho^{12}$, $\rho^{21}$, and $\rho^{22}$. 
The result is 
\begin{align}
\rho^{11}&=\dfrac{-\Gamma_{12}+2\Im{G^*F}}{4\Im{G^*F}-\gamma},\no\\
\rho^{12}&=\dfrac{(\Gamma_{12}-\Gamma_{21})}{4\Im{G^*F}-\gamma}F={\rho^{21}}^*,\no\\
\rho^{22}&=\dfrac{-\Gamma_{21} +2\Im{G^*F}}{4\Im{G^*F}-\gamma},
\end{align}
where $\gamma=2\gamma_{12}$ and we have defined 
\begin{align}
F=&-iG/J, \no\\
G=&\bra{E_1}H_0^{(0)}\ket{E_2}+\bra{E_1}\Delta H_{\rm{FE}}^{(1)}\ket{E_2},\no\\
J=&i(\bra{E_1}H_0^{(0)}+\Delta H_{\rm{FE}}^{(1)}\ket{E_1}-\bra{E_2}H_0^{(0)}+\Delta H_{\rm{FE}}^{(1)}\ket{E_2})\no\\
&+\gamma/2.
\end{align}

To obtain an explicit form of the density matrix in the high-frequency regime, we 
expand $F$ and $\Im{G^*F}$ in terms of $1/\omega$ and we define 
\begin{align}
F=F^{(0)}+F^{(1)}+F^{(2)}+\cdots,\no\\
\Im{G^*F}=\mathcal{F}^{(0)}
+\mathcal{F}^{(1)}+\mathcal{F}^{(2)}+\cdots,
\end{align}
where $F^{(m)}$ and $\mathcal{F}^{(m)}$ are respectively the $(1/\omega)^m$-order terms of $F$ and $\Im{G^*F}$. As a result, the density matrix in the NESS is given by 
\begin{align}
\rho^{11}&=\frac{1}{\lambda}(\Gamma_{12}-2(\mathcal{F}^{(0)}+\mathcal{F}^{(1)}))\no\\
&+\frac{4}{\lambda^2}(\Gamma_{12}-2\mathcal{F}^{(0)})4\mathcal{F}^{(1)}+\order{\omega^{-2}},\no\\
\rho^{12}&=(\Gamma_{12}-\Gamma_{21})
\Big(-(F^{(0)}+F^{(1)})/\lambda-4F^{(0)}\mathcal{F}^{(1)}/\lambda^2
\Big) \no\\
&+\order{\omega^{-2}}={\rho^{21}}^*,\no\\
\rho^{22} &=\frac{1}{\lambda}(\Gamma_{21}-2(\mathcal{F}^{(0)}+\mathcal{F}^{(1)}))\no\\
&+\frac{4}{\lambda^2}(\Gamma_{21}-2\mathcal{F}^{(0)})\mathcal{F}^{(1)}+\order{\omega^{-2}},
\label{rhosNESS}
\end{align}
where we have introduced new parameters 
\begin{align}
\lambda&=\frac{2\gamma}{\kappa}|\bra{E_2}H_0^{(0)}\ket{E_1}|^2+\gamma,\no\\
\kappa&=(\bra{E_1}H_0^{(0)}\ket{E_1}-\bra{E_2}H_0^{(0)}\ket{E_2})^2+(\gamma/2)^2.
\end{align}

At the end of the subsection, 
we comment on a simple case of $H_0^{(0)}=H_{\rm stat}$. 
In this case, the computation flow of Appendix~\ref{sec:A-SolNESS-eq} is still applicable if $H_0$ and $\Delta H_{\rm FE}$ are respectively replaced with $H_0^{(0)}$ and $\Delta H_{\rm FE}+H_0^{(1)}$.


\subsection{\label{sec:A-SolNESS-propo}\texorpdfstring{$H_0=CH_{\rm stat}$}{H0=CH}}
Here, we shortly consider a special case of $H_0\neq H_{\rm stat}$ in two-level systems. Namely, we consider the case where $H_0$ is proportional to $H_{\rm stat}$: $H_0=CH_{\rm stat}$ with $C$ being a constant independent of $\omega$.  
In this case, $\bra{E_n}H_0\ket{E_m}=C\delta_{nm}E_m$ holds and it leads to $\lambda=\gamma$, 
$\mathcal{F}^{(0)}=\mathcal{F}^{(1)}=0$, and $F=\frac{\bra{E_1}\Delta H_{\mr{FE}}\ket{E_2}}{-C(E_1-E_2)+i\gamma/2}
+\order{\omega^{-2}}$. 
Hence, the diagonal components of the density matrix $\tilde \rho_\infty$ are 
\begin{align}
\rho^{kk}&=\rho^{k}_{\mathrm{\mr{can}}}+\order{\omega^{-2}}=
\frac{e^{-\beta E_k}}
{\sum_{l}e^{-\beta E_l}}+\order{\omega^{-2}}.
\label{eq:diagonalCH}
\end{align}
The off-diagonal components $\rho^{kl}\;(k\neq l)$ are
\begin{align}
\label{eq:offdiagonalCH}
\rho^{kl}&=\frac{ \bra{E_k}\Delta H_{\mr{FE}}\ket{E_l}}{C(E_k-E_l)-i\gamma/2}(\rho^{k}_{\mathrm{\mr{can}}}-\rho^{l}_{\mathrm{\mr{can}}})+\order{\omega^{-2}}.
\end{align}
Equations~(\ref{eq:diagonalCH}) and (\ref{eq:offdiagonalCH}) are still applicable in generic multi-level systems if we replace $\gamma/2$ with $\gamma_{kl}$ in Eq.~(\ref{eq:offdiagonalCH}).

\subsection{\label{sec:A-SolNESS-Lz}\texorpdfstring{$\Gamma_{ii}\neq 0$}{Gamma{ii} not equal 0}}
Finally, we consider the case that a ``diagonal'' jump operator $L_{ii}=|E_i\rangle\langle E_i|$ exists under the condition of $H_0\neq H_{\rm stat}$. For simplicity, we focus on two-level systems. As in Appendix~\ref{sec:A-GKSLBloch-Lz}, 
the diagonal jump operator is given by
\begin{align}
    L_{3}=\sqrt{\frac{\gamma_\perp}{2}}\,\,\sigma_{z}.
\end{align}
For this setup, the dissipation term of the GKSL equation is written as 
\begin{align}
\label{eq:GKSL_d}
 \mathcal{D}(\rho)=\sum_{i,j (i\neq j)}\Gamma_{ij}\Big[L_{ij}\rho L_{ij }^\dagger
-\frac{1}{2}\qty{L_{ij}^\dagger L_{ij},\rho}\Big]\no\\
+\Big[L_{3}\rho L_{3}^\dagger
-\frac{1}{2}\qty{L_{3}^\dagger L_{3},\rho}\Big],
\end{align}
where off-diagonal jump operators $L_{12,21}$ are assumed to satisfy the detailed balance condition of Eq.~(\ref{eq:Gamma}). Computing the matrix elements $\bra{E_n}\LL_{\rm{eff}}\tilde\rho_{\infty}\ket{E_m}$ with the dissipation term of Eq.~(\ref{eq:GKSL_d}), we can obtain the generic formula for the density matrix in the NESS. 
The result is almost the same as Eq.~(\ref{rhosNESS}), but we should respectively replace the parameters $\lambda$ and $\kappa$ with 
\begin{align}
\tilde{\lambda}&=\frac{2}{\tilde\kappa}(\gamma+2\gamma_\perp)
|\bra{E_2}H_0^{(0)}\ket{E_1}|^2
+{\gamma},\no\\
\tilde\kappa&=(\bra{E_1}H_0^{(0)}\ket{E_1}-\bra{E_2}H_0^{(0)}\ket{E_2})^2+(\gamma+2\gamma_\perp)^2/4.
\end{align}
The generalization to multi-level systems is straightforward.

\section{\label{App:numerical} Additional results of pulse-driven precession}
In Sec.~\ref{sec:Pulse-Prece}, we have analyzed the laser-pulse driven spin dynamics in a ferromagnetic metal state with magnetization being a moderate value ($S_{\rm occ}^x=0.248$). As we mentioned, the reason why we choose a moderate value $S_{\rm occ}^x=0.248$ is that in real metallic magnets, only a part of conducting electrons near Fermi surface contribute to the magnetic order~\cite{white2007}. 
In this section, aside from such a realistic setup, 
we show the numerical results of spin dynamics in two extreme cases: The first case is the paramagnetic metal state without mean field ($B_x=0$) and the second is a nearly saturated state with $S_{\rm occ}^x\to +0.5$.

\begin{figure*}[t]
\includegraphics[width=0.95\textwidth]{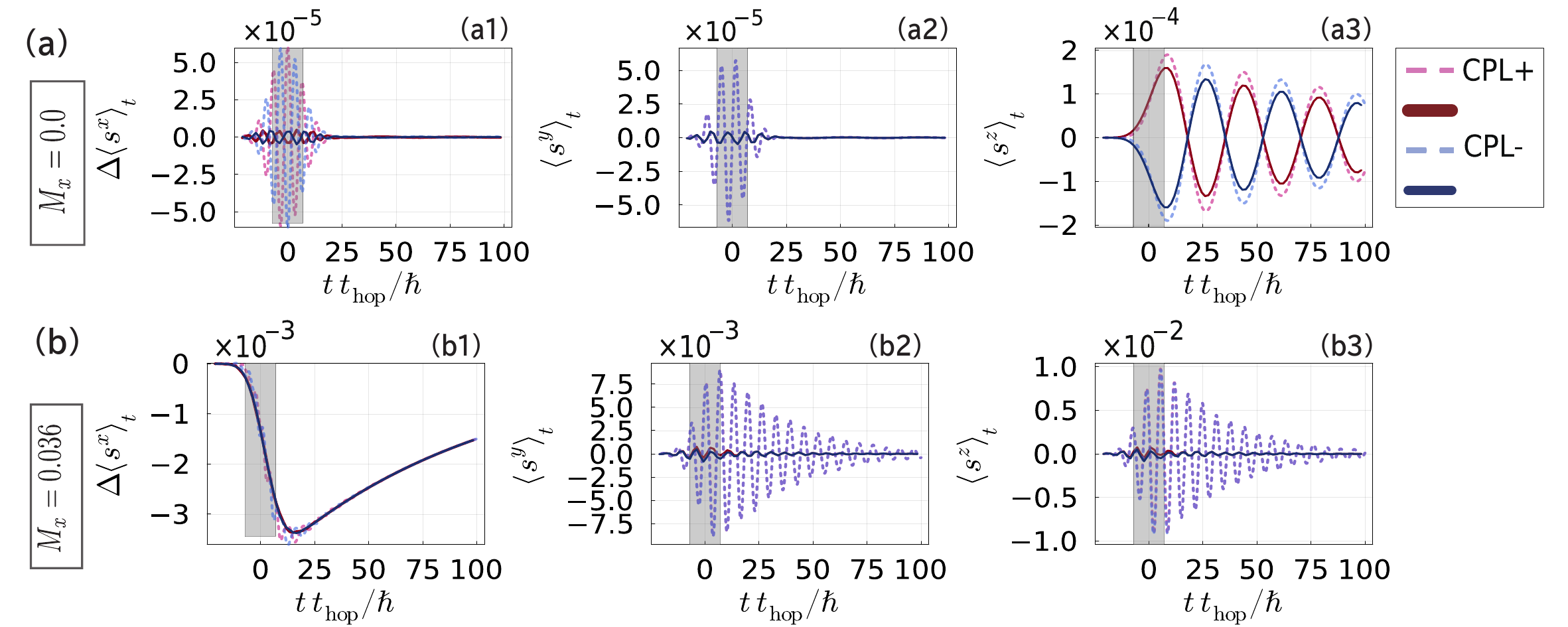}    
\caption{
Time evolution of $\ev {s^x}_t$, $\ev {s^y}_t$, and $\ev {s^z}_t$ computed by the GKSL equation for (a) a paramagnetic metal state ($S_{\rm occ}^x=0$ and $B_x=0$) and (b) a nearly saturated state ($S_{\rm occ}^x\simeq 0.494$) in the model~(\ref{Eq:timedepH_pulse}) at $T=0$. Dotted lines denote $\ev {s^\alpha}_t$, while solid lines are the slow modes defined by $s_{\rm slow}^\alpha(t)=\frac{1}{2T_\omega}\int_{t-T_\omega}^{t+T_\omega} dt' \Delta\langle s^\alpha\rangle_{t'}$. Symbols CPL$+$ and CPL$-$ respectively correspond to right ($\hbar \omega/t_{\mr{hop}}=1$) and left ($\hbar \omega/t_{\mr{hop}}=-1$) circularly polarized pulses. The gray region stands for the width of the laser pulse $\tau$. 
Other parameters are all the same as those of Fig.~\ref{PulseMxLowhigh}(a): 
$t_{\rm hop}=1$, $eaE_0/t_{\mr{hop}}=0.5$, $\alpha/t_{\mr{hop}}=0.1$, and $\gamma/t_{\mr{hop}}=0.01$.
}
\label{fig:Precession_2}
\end{figure*}

From Fig.~\ref{fig:Precession_2}(a), one sees that there is no pulse-driven precession of the $y$ component of electron spins in the paramagnetic state. This is a natural result because we have no magnetic moment unlike Fig.~\ref{Fig:PulsePonchi}. Instead, we find that the laser pulse induces a small magnetization along $S^z$ axis and it may be interpreted as a short-time version of IFE in the NESS. 
Figure~\ref{fig:Precession_2}(b) shows that in the nearly saturated state, the $y$ and $z$ components of spins have only a very fast oscillation, whose frequency is the same as the laser one $\omega$. 
The numerical result indicates that FE does not occur well in this state. This would be because the direction of spin moment is strongly locked by a strong mean field $B_x$ and laser cannot change their direction and magnitude well. 

From these results, we can conclude that the ferromagnetic metal state with a small or moderate magnetization, that we have used in Sec.~\ref{sec:Pulse}, is close to a real setup of IFE in magnetic systems.  

%

\end{document}